\documentclass[10pt]{article}

\usepackage{etex}
\usepackage[T1]{fontenc}
\usepackage{mathtools}
\usepackage{amsmath}
\usepackage{amssymb}
\usepackage{esint}
\usepackage{caption}
\usepackage{subfig}

\usepackage[pdf]{pstricks}
\usepackage{pst-func}
\usepackage{pst-math}
\usepackage{pstricks-add}
\usepackage{pst-solides3d}

\usepackage{tikz}
\usetikzlibrary{positioning}
\usetikzlibrary{through,backgrounds}
\usetikzlibrary{calc,fadings,decorations.pathreplacing}
\usetikzlibrary{intersections}

\newcommand{\pdd}[2]{\frac{\partial #1}{\partial #2}}
\newcommand{\dd}[2]{\frac{d #1}{d #2}}
\newcommand{\pd}[1]{\frac{\partial}{\partial #1}}
\renewcommand{\d}[1]{\frac{d}{d#1}}

\renewcommand{\div}[1]{\nabla\cdot #1}
\newcommand{\grad}[1]{\nabla #1}

\newcommand{\beq}{\begin{equation}}
\newcommand{\eeq}{\end{equation}}

\newcommand{\secn}[1]{section-\ref{#1}}

\newcommand{\fig}[1]{fig.-\ref{#1}}
\newcommand{\Fig}[1]{Fig.-\ref{#1}}

\newcommand{\eq}[1]{eq.-\ref{#1}}
\newcommand{\Eq}[1]{Eq.-\ref{#1}}

\newcommand{\dx}{\Delta x}

\newcommand\pgfmathsinandcos[3]{%
  \pgfmathsetmacro#1{sin(#3)}%
  \pgfmathsetmacro#2{cos(#3)}%
}
\newcommand\LongitudePlane[3][current plane]{%
  \pgfmathsinandcos\sinEl\cosEl{#2} 
  \pgfmathsinandcos\sint\cost{#3} 
  \tikzset{#1/.estyle={cm={\cost,\sint*\sinEl,0,\cosEl,(0,0)}}}
}
\newcommand\LatitudePlane[3][current plane]{%
  \pgfmathsinandcos\sinEl\cosEl{#2} 
  \pgfmathsinandcos\sint\cost{#3} 
  \pgfmathsetmacro\yshift{\cosEl*\sint}
  \tikzset{#1/.estyle={cm={\cost,0,0,\cost*\sinEl,(0,\yshift)}}} %
}
\newcommand\DrawLongitudeCircle[2][1]{
  \LongitudePlane{\angEl}{#2}
  \tikzset{current plane/.prefix style={scale=#1}}
  \pgfmathsetmacro\angVis{atan(sin(#2)*cos(\angEl)/sin(\angEl))} %
  \draw[current plane,ultra thick] (\angVis:1) arc (\angVis:\angVis+180:1);
  \draw[current plane,dashed,ultra thick] (\angVis-180:1) arc (\angVis-180:\angVis:1);
}
\newcommand\DrawLatitudeCircle[2][1]{
  \LatitudePlane{\angEl}{#2}
  \tikzset{current plane/.prefix style={scale=#1}}
  \pgfmathsetmacro\sinVis{sin(#2)/cos(#2)*sin(\angEl)/cos(\angEl)}
  \pgfmathsetmacro\angVis{asin(min(1,max(\sinVis,-1)))}
  \draw[current plane,ultra thick] (\angVis:1) arc (\angVis:-\angVis-180:1);
  \draw[current plane,dashed,ultra thick] (180-\angVis:1) arc (180-\angVis:\angVis:1);
}

\pgfdeclareradialshading{ballshading}{\pgfpoint{-10bp}{10bp}}
 {color(0bp)=(gray!10!white); 
 color(12bp)=(gray!20!white);
 color(30bp)=(gray!30!gray); 
 color(40bp)=(gray!40!gray); 
 color(50bp)=(gray)}
 
\tikzset{%
  >=latex, 
  inner sep=0pt,%
  outer sep=2pt,%
  mark coordinate/.style={inner sep=0pt,outer sep=0pt,minimum size=3pt,
    fill=black,circle}%
}

\begin{document}

\title{Continuity, the Bloch-Torrey equation, and diffusion MRI}

\author{Matt G Hall}

\maketitle

\abstract{
The Bloch equation describes the evolution of classical particles tagged with a magnetisation vector in a strong magnetic field and is fundamental to many NMR and MRI contrast methods. The equation can be generalised to include the effects of spin motion by including a spin flux, which typically contains a Fickian diffusive term and/or a coherent velocity term. This form is known as the Bloch-Torrey equation, and is fundamental to MR modalities which are sensitive to spin dynamics such as diffusion MRI. Such modalities have received a great deal of interest in the research literature over the last few years, resulting in a huge range of models and methods. 

In this work we make make use of a more general Bloch-Torrey equation with a generalised flux term. We show that many commonly employed approaches in Diffusion MRI may be viewed as different choices for the flux terms in this equation. This viewpoint, although obvious theoretically, is not usually emphasised in the diffusion MR literature and points to interesting new directions for methods development.

In the interests of completeness, we describe all the necessary mathematical and physical framework -- including the method used to solve the various Bloch-Torrey equations. We then review the relevant portions of the diffusion MRI literature and show how different flux choices lead to different models. All the models considered here are continuum-based but we also discuss their relationship to so-called microstructure methods and other more advanced ideas such as time-dependent diffusivities.
}
\\
\\
\textbf{Keywords}: Diffusion MRI, HARDI, Diffusion Tensor Imaging, Diffusion Kurtosis Imaging, Microstructure Imaging, Continuity Equation

\section{Introduction}
The manipulation of spin magnetic moments is central to NMR and MRI. Via a wide variety of mechanisms, it is possible to excite signals of various kinds from samples placed in the scanner's magnetic field by applying controlled RF pulses and magnetic field gradients. Since the behaviour of spins in affected by the chemical and physical environment they experience, NMR and MRI measurements provide a vector for analysing the chemical make-up and physical structure of objects or living organisms. 

A classical description of spin magnetisation in a strong magnetic field is provided by the Bloch equation. This has the form
\beq
    \dd{\mathbf{M}}{t}= \gamma \mathbf{M}\times\mathbf{B} - \frac{M_x\mathbf{i} + M_y\mathbf{j}}{T_2} - \frac{M_z-M_0}{T_1}\mathbf{k}
    \label{eq:bloch}
\eeq
where $\mathbf{M}=(M_x,M_y,M_z)^T$ is the local magnetisation, $M_0$ is the equilibrium magnetisation, $\mathbf{B}$ is the static scanner field, $T_1$ and $T_2$ are relaxation constants and $\gamma$ is the gyromagnetic constant for the medium. We have also made use of $\mathbf{i}$, $\mathbf{j}$, and $\mathbf{k}$ as the unit vectors defining the cardinal direction in the lab frame.

The Bloch equation describes the change in magnetisation of a continuum of spins in a field $\mathbf{B}$, assuming that spins are stationary. It is an effective theory which captures smaller-scale effects like spin-spin and spin-lattice interaction via decay constants. In many situations, however, it is useful to consider the effects of diffusive motion on spin magnetisation, which can have an important effect on NMR measurements. Spin diffusion was first incorporated by Torrey \cite{torrey}, who included an additional additive diffusion term. Torrey's original formulation may be written as
\beq
    \dd{\mathbf{M}}{t}= \gamma \mathbf{M}\times\mathbf{B} - \frac{M_x\mathbf{i} + M_y\mathbf{j}}{T_2} - \frac{M_z-M_0}{T_1}\mathbf{k} - \div{(D\grad{\mathbf{M}})}
    \label{eq:bt}
\eeq
where the new term reduces to $D\nabla^2\mathbf{M}$ when $D$ is constant and reduces to the Bloch form in the limit of zero diffusion current. Torrey's derivation of this form assumes that spins have only a very small drift velocity, and in more recent literature it is common to also include a linear flow term defined by the vector $\mathbf{u}$ which gives 
\beq
\dd{\mathbf{M}}{t}= \gamma \mathbf{M}\times\mathbf{B} - \frac{M_x\mathbf{i} + M_y\mathbf{j}}{T_2} - \frac{M_z-M_0}{T_1}\mathbf{k} - \div{(D\grad{\mathbf{M}})} - \div{\mathbf{uM}.}
    \label{eq:bt_flow}
\eeq
The Bloch-Torrey equation lead to the development of various pulse sequences which allow the diffusive term to be quantified (see, e.g \cite{CarrPurcell}), which in turn has lead to the development of diffusion-weighted imaging (DWI).

Diffusive motion encodes information about the environment experienced by diffusing particles. This encoding, however, is non-trivial and extracting environmental information from measurements of diffusion, particularly when it happens in the presence of microstructure, is hugely challenging. Nevertheless, the fact that the length scales of diffusive motion over the timescale of a typical MR pulse sequence are orders of magnitude smaller than a  typical scan voxel and this has lead to considerable research effort into how best to analyse diffusion-weighted measurements. This in turn has lead to a large number of models and approaches to diffusion imaging which can be confusing to someone new to the field.

All of diffusion imaging is ultimately grounded in the Bloch-Torrey equation -- different solutions provide the models used to analyse diffusion-weighted data. This paper presents a derivation of the Bloch-Torrey equation which emphasises a general transport process. This generalised form allows a great deal of the modelling work in diffusion imaging to be combined within a single framework, and it makes clear the relationships between approaches which can otherwise seem highly disparate and arbitrary. We emphasise, however, that there is no suggestion that we should somehow optimise a flux process to explain data, just that by making a choice of model to fit to data is in many cases equivalent to making a choice of flux.

The Bloch-Torrey equation, in either of the above forms, implicitly treats magnetisation as a continuum. The presence of the spatial derivative requires that $\mathbf{M}$ be smoothly varying in space. Similarly, the time derivative assumes that the continuum is changing smoothly with time. This is compatible with the concept of magnetisation as a vector field rather than being defined discretely on separate, point-like particles. We can think of this as a ``local magnetisation'' -- the mean magnetisation per unit volume -- where
\beq
\mathbf{M} = \frac{1}{V}\sum_{i=1}^{N}\vec{\mathbf{\mu}}_i
    \label{eq:mag_dens}
\eeq
where $\vec{\mathbf{\mu}}_i$ are magentisations due to point charges in the applied field $\mathbf{B}$ and $V$ is a small control volume over which the derivatives are smooth. 

Another important point mathematically is that because the derivatives operate on a vector quantity, their meaning is a little different to that for scalar quantities. The Bloch-Torrey equation contains the gradient operator acting on a vector quantity, for example, and \eq{eq:bt_flow} contains a product between the vector $\mathbf{u}$ and the magnetisation. Some care is required to ensure that the rank or the resulting objects is handled correctly and that differential operators are correctly defined in this context. An explanation of how this is handled requires new mathematical tools and definitions.

This paper derives of the Bloch-Torrey equation in terms of the continuity equation for a vector quantity. It aims to be as self-contained as possible, covering the necessary mathematical preliminaries before deriving the equation and discussing interpretation of the various terms. In particular we will cover the Dyadic Product and Tensor contraction, which we need to examine the rank of the vector equations we will derive. We will also consider affine transforms and homogeneous coordinates, which will be useful to streamline our notation and make the behaviour described by the equation clearer. We then introduce the theoretical concepts we will make use of, such as the continuity equation, and derive the Bloch-Torrey equation. Finally, we will solve this equation employing different theoretical assumptions and review the diffusion MRI modelling literature. The aim is to give a comprehensive overview of the diffusion MRI modelling literature whilst also providing a strong theoretical foundation. 

In each case we will show that different models are solutions to different forms of the generalised equation derived here, and thus corresponds to an implicit set of theoretical assumptions. The approach we employ makes the relationships between different models apparent and also points towards new research directions for models development which we will discuss in the final section of the paper.

\section{Mathematical preliminaries} 
This section introduces some simple but non-standard mathematics necessary for the rest of the paper. The emphasis here will be to give enough of a background to enable the interested reader to follow the arguments in the remainder of the paper. As such we will provide only an introduction and point to references where further detail can be found.
\subsection{The Dyadic product}
\label{dyad-prod}
Products of vectors are familiar from high-school and undergraduate applied maths courses. These courses tell us that, unlike scalars, there is more than one way multiply pairs to vectors. We can combine the vectors to form a scalar (often called the dot, or inner, product), an operation that reduces the ranks of the object by one (vector to scalar), and we can combine vectors to form another vector (the cross product) which leaves rank unchanged. A related concept is the multiplication of a vector by a scalar, which also yields a vector -- one with altered length but direction preserved from the vector multiplicand.

Given the above we might ask whether there is an operation that, for example, raises the rank of the tensors, a ``tensor product'' or similar. The answer, as it turns out is yes. In fact, there are several related products that do this, but for our purposes the most convenient is the \emph{Dyadic or Outer product}, named because it is constructed from products of pairs of elements of the constituent vectors, known as dyads. The dyadic product was originally proposed by Gibbs in 1884 \cite{gibbsdyad1884}, and is used widely in the theory of continuum mechanics, although in more modern literature outer product is a more common term.

The dyadic product in Euclidean space is relatively simple to construct. We define the product $\mathbf{a}\mathbf{b}$ as a matrix in which each element $M_{ij}$ is the product of the $i$th and $j$th elements of the two constituent vectors. I.e.
\beq
	\mathbf{a}\mathbf{b} = \left(\begin{array}{ccc}
								a_1b_1 & a_1b_2 &a_1b_3 \\
								a_2b_1 & a_2b_2 &a_2b_3 \\
								a_3b_1 & a_3b_2 &a_3b_3 \\
							\end{array}\right),
\eeq
which is formally equivalent to the product of a column vector with a row vector -- $\mathbf{a}\mathbf{b} = \mathbf{a}\mathbf{b}^T$. The notation $\mathbf{ab}$ was established by Gibbs \cite{gibbsdyad1884} and is commonly used in the fluid dynamics literature. We will use this notation here.

Conceptually, we can think of the outer product as a set of projections. The first column of the matrix is the product of the vector $\mathbf{a}$ with the (scalar) first element of the vector $\mathbf{b}$ -- a vector with direction parallel to $\mathbf{a}$ with length $|\mathbf{a}|b_1$. Compare this to the definition of momentum, which is a vector quantity with direction defined by velocity and length defined by the magnitude of velocity and a scalar quantity, mass. A similar analogy holds for the second and third columns of the matrix, which shows that the outer product is an extension of the idea of the multiplication of a vector by a scalar -- the matrix describes the effect of scaling one vector by another. This interpretation is particularly useful in the case where the two vectors span different spaces. In our current context, one vector represents quantity associated with real space, such as position or velocity, whereas the other represent magnetisation. The dyadic product describes a projection from one space into another. This turns out to be a useful concept in the continuity approach we will employ, especially in situations where we can think of particles and continua that are ``tagged'' with some quantity of interest.
 
\subsection{Tensor contraction}
\label{tens-contr}
In the previous section we saw how a product operation can leads to a increase in rank in the resulting object. We already know that there are other operations which lower an objects rank, and in this section we discuss a new one. Specifically, we are interested in how the divergence theorem generalises vector objects to higher rank, tensor objects.

The divergence relation itself is fundamental to vector calculus. It relates the flux of a quantity through a closed surface to the behaviour of the quantity in the enclosing volume - it is a multidimensional generalisation of the fundamental theorem of calculus. The theorem simply states that the integral of the component of the flux of some vector quantity normal to a closed surface over the surface is equal to the volume integral of the divergence of the quantity over the enclosed volume,
\beq
	\oiint_S \mathbf{M}\cdot dS = \iiint_V \div{\mathbf{M}}dV.
	\label{divthm}
\eeq
This is valid in any number of dimensions, although for historical reasons the names of the theorems differ in one (Green's Theorem), two (Gauss's Theorem or Divergence Theorem), and higher dimensions (Stokes' Theorem).

This can also be viewed as a relationship in linear algebra, and we might ask how it would generalise to objects with higher rank than vectors. If we re-write \eq{divthm} in Einstein notation\footnote{i.e. summing over repeated indices} we get
\beq
	\oiint_S \mathbf{M}_in_idS = \iiint_V \pdd{\mathbf{M}_i}{x_i}dV
\eeq
which immediately suggests we can replace the vector with a higher rank object $M_{i_1i_2\ldots i_n}$to give
\beq
	\oiint_S M_{i_1i_2\ldots i_n} n_{i_q}dS = \iiint_V\pdd{M_{i_1i_2\ldots i_n}}{x_{i_q}}dV.
\eeq
Where we're choosing a particular index to contract $i_q$ by summation \cite{RileyHobson2010}. In our case the maximum rank we will encounter is two. In this case the above reduces to 
\beq
	\oiint_S M_{ij} n_{k}dS = \iiint_V\pdd{M_{ij}}{x_{k}}dV,
    \label{eq:tenscontr}
\eeq
where $k\in\{i,j\}$.

\Eq{eq:tenscontr} is known as the \emph{tensor contraction}. In common with familiar notions like the trace of a matrix, it provides a measure of a tensor quantity which is independent of the coordinate system employed. It is widely employed in differential geometry where it can be used to change the basis used to define a quantity or to convert between covariant and contravariant quantities. The important aspect of tensor contraction in the current context is that it lowers the rank of an object.

\subsection{Homogeneous coordinates and affine transforms}
\label{homog}
This section describes homogeneous coordinates. This is a system of coordinates originally developed in projective geometry \cite{homog_history} where they are useful because in addition to the regular points of Euclidean space, they also include so-called ``points at infinity'', defined as the limiting position of a point moving in a given direction. Homogeneous coordinates often simplify relationships in projective geometry, and for our purposes their most useful property is allowing affine transforms to be written as simple matrix multiplications. This section describes homogeneous coordinates, affine transformations, and how the two are related.

An affine transformation defines a relationship between two spaces which preserves points, lines, and planes -- the affine transform of a point is another point, that of a line is another line, etc. The group of affine transforms includes any transform which preserves points, straight lines, and planes: rotations, scaling transforms, sheers, reflections,  and translations. Transformations which do not preserve these objects, such as a non-linear warp are not part of the group of transformations. The non-linear warp transforms straight lines to curves, for example. Affine transformations are fundamental objects in computer graphics, where they are used to define relationships between objects in a scene to be rendered, and are also useful for our current purposes. 

Although it is common to approach affine transformations from an abstract, group-theoretic perspective it is more pertinent here to consider how they can be represented. One interesting aspect of these transformations is that while rotations, sheers, reflections and scalings are all multiplicative, translations are additive. This complicates representation of a general affine transform as it must include both multiplicative and additive terms. In general, the affine transform of a vector $\mathbf{x}\to \mathbf{y}$ may be written as
\beq
	\mathbf{y}=A\mathbf{x}+\mathbf{x}_0
\eeq
where $A$ is a matrix describing the multiplicative transform and $\mathbf{x}_0$ is a vector describing the additive part. Although this has the form of a linear transform, it is possible to simplify this to a single matrix multiplication via the use of an \emph{augmented matrix} by noting that
\beq
	\left(\begin{array}{c}
		\mathbf{y}\\
		1\\
	\end{array}\right) = \left(\begin{array}{cc}
					A & \mathbf{x}_0 \\
					0, \ldots, 0 & 1 \\
				\end{array}\right)
				\left(\begin{array}{c}
					\mathbf{x}\\
					1\\
				\end{array}\right).
\eeq
By adding an extra coordinate to the vector we introduce a trivial additional equation ($1=1$) and are able to include the additive part of the transform into the matrix. Aside from neatness, this also gives use a more direct way to solve the equation in terms of the eigensystem of the matrix.

The question remains, however, of why we would want to introduce this extra coordinate and what it might mean in terms of the original coordinate space. What would it mean if its value were not unity, for example? This is the domain of homogeneous coordinates.

Homogeneous coordinates were first introduced as a tool for studying projective geometry by M\"obius in $1827$ \cite{mobius}. The motivation was to extend Euclidean coordinates to include the aforementioned points at infinity whilst simplifying the algebra of projective spaces. The idea is to establish a dual relationship between lines and points by adding an additional coordinate to the space. A point in 3D space may be represented by four coordinates such that
\beq
	(x,y,z) \to (X,Y,Z,W)
\eeq
where the $W$ coordinate is interpreted as a multiplying factor on the other three. Specifically,
\beq
	(x,y,z) = (\frac{X}{W},\frac{Y}{W},\frac{Z}{W},\frac{W}{W}) =(\frac{X}{W},\frac{Y}{W},\frac{Z}{W},1).
\eeq
The presence of the $W$ coordinate acts like an overall scale on the other coordinates. Changing the value of $W$ uniformly changes the magnitude of the other coordinates, but does not change the overall spatial relationships between points -- increasing $W$ decreases the overall scale, decreasing $W$ increases it. This means that if we define a shape in homogeneous coordinates, changing the value of $W$ will make it bigger or smaller but not fundamentally alter the spatial relationships within it. 

This ``homogeneity'' of spatial relationships with respect to changes in the additional coordinate is what gives homogeneous coordinates their name. A good analogy to this property is an image projected onto a screen -- moving the projector back or forth will change the size of the image on the screen but not the image itself. An additional property is that homogeneous coordinates also include so-called points at infinity. These correspond to a choice of direction, and occur when $W=0$ correspond to ``points at infinity'', which are points are unaltered by translations.

Notice how well this compliments the augmented matrix notation from affine transforms: homogeneous coordinates provide both the rationale behind and the interpretation of the additional coordinate in the augmented affine transform matrix. We will see in \secn{blochterms} the this provides a simple and compact representation of the Bloch equation which also enables it to be solved in a a powerful and succinct way.

\subsection{Matrix differential equations}
\label{matrix-eqs}
Magnetisation is a vector quantity, at least in the classical description commonly employed in MRI (see \eq{eq:mag_dens}). As such the evolution equations describing it are expressed in terms of vectors with matrix coefficients, and in most cases contain only a first-order time derivative. The solutions of these equations describe the behaviour of spins in magnetic fields, and are used to analyse data measured from MR experiments. Later in this work we will show how different choices lead to different models used in diffusion MRI and beyond. We want to be able to solve the multi-variable partial differential equations as complete entities rather than take individual component equations individually. This section outlines that methodology. 

\subsubsection{The form of the equation to solve}
Differential equations in vector quantities employ matrices as coefficients. This is helpful from the point of view of solving them, but leads to a confusion in terminology -- they are known both as vector differential equations \cite{jacobs2010} and as matrix differential equations \cite{MoyaCessa2011}. We'll use the former these here. Vector differential equations often take the following form
\beq
    \pdd{}{t}\mathbf{M}=\left(\begin{array}{c}\pdd{M_x}{t}\\\pdd{M_y}{t}\\\pdd{M_z}{t}\\\end{array}\right)=A\mathbf{M}
    \label{eq:vecdiff}
\eeq
where $A$ is a matrix describing the dependencies between the different elements of the vector $\mathbf{M}$. As it stands, \eq{eq:vecdiff} only contains a time derivative. With the sole exception of the Bloch equation, all the equations we are interested contain at least one spatial derivative and so take the form
\beq
    \pdd{\mathbf{M}}{t}=A\mathbf{M} + \div{J(\mathbf{M})}
    \label{eq:vecdiff2}
\eeq
where $\mathbf{J}$ is a known time-independent function. We can reduce this to the form of \eq{eq:vecdiff2} by taking the spatial Fourier transform 
\beq
    f(\mathbf{q})= \int F(\mathbf{x})e^{-\mathbf{q}\cdot\mathbf{x}}d\mathbf{x}.
    \label{eq:ft_q}
\eeq
This gives
\beq
    \pdd{\mathbf{m}}{t}=A\mathbf{m} -i\mathbf{q}\cdot j(\mathbf{m})
    \label{eq:vecdiff_ft}
\eeq
where $j$ and $\mathbf{m}$ are the spatial Fourier transforms of $J$ and $\mathbf{M}$ respectively, $\mathbf{q}$ is the Fourier variable and $i$ is the square root of -1. Assuming the second RHS term is proportional to $\mathbf{m}$ we can factor out the magnetisation such that $j=\mathbf{jm}$ to give
\beq
    \pdd{\mathbf{m}}{t}=A\mathbf{m} -i\mathbf{q}\cdot \mathbf{j}\mathbf{m}.
\eeq
The dot product in the last term yields a scalar, which is equivalent to multiplication by a diagonal matrix $I(\mathbf{q}\cdot\mathbf{j})$ where $I$ is the identity matrix. In this form \eq{eq:vecdiff_ft} becomes
\beq
    \pdd{\mathbf{m}}{t}=(A-i(\mathbf{q}\cdot\mathbf{j})I)\mathbf{m},
    \label{vecdiff_combo}
\eeq
which is the same form as \eq{eq:vecdiff}. This is a useful property since it means we can apply the techniques for solving ordinary vector differential equations to the partial vector differential equations we are interested in. Notice also that each matrix that contributes to $A$ also makes an additive contribution to the Eigen\emph{values} of $A$. 

\subsection{The Fourier transform and q-space formalism}
At this point it is worth discussing the above procedure in light of diffusion MR formalism and assumptions. \Eq{eq:ft_q} defines the spatial Fourier Transform of the spatial coordinate. Diffusion MRI contains more than one concept of a spatial Fourier transform and both are used to construct images and analyse data. The first of these concepts is the relationship between image data measured by the scanner and the images constructed. This relationship is via the spatial Fourier transform of \emph{position} to spatial frequency. MR measurements are in the frequency domain, images in the position domain. The frequency domain is known as $\mathbf{k}$-space, and is fundamental to all MR imaging. 

Diffusion MRI also involves the concept of $\mathbf{q}$-space \cite{callaghan}\cite{callaghan-qspace}\cite{cory}. $\mathbf{q}$-space shares a Fourier relationship with the space of diffusion displacements measured during the pulse sequence provided that the diffusion-weighting gradient pulses are very short compared to the diffusion time. Under these circumstances $\mathbf{q}$-space and diffusion displacement space are connected via a Fourier transform, and we can define
\beq
    \mathbf{q}= \gamma\delta\mathbf{G}.
    \label{eq:qdef}
\eeq

The choice of notation of $\mathbf{q}$ in the Fourier approach detailed above is deliberate -- it is intended to denote the $\mathbf{q}$-space approximation. The Bloch-Torrey equation is formulated with the spatial variable corresponding to the diffusion process, not the spatial position in an image, and thus $\mathbf{q}$-space is a more appropriate interpretation of the Fourier variable. In some definitions in the literature, $\mathbf{q}$ contains an additional factor of $2\pi$. This is dependent on the units of $\gamma$, converting cycles per second to radians per second or vice versa. We have chosen to omit this factor here for brevity.

In practical terms the two transforms are related as follows. A diffusion-weighted acquisition is made -- in general this requires more than one image to be acquired (an ADC map requires a minimum of two images: one weighted, one unweighted \cite{ADC}. DTI requires a minimum of seven images \cite{basser1994}). Each of these is acquired in $\mathbf{k}$-space and transformed into image space to allow the image to be human-interpretable. The set of images is then co-registered into the same space, which aligns the voxels. The same voxel in each image represents a set of measurements in $\mathbf{q}$-space, which can be analysed by various choices of model. Since the domain of diffusion-weighted measurements is $\mathbf{q}$-space, and it is here where models are typically formulated.

An alternate approach is to transform the q-space data into diffusion-displacement space directly -- this is known as q-space imaging \cite{callaghan} and requires that Fourier relationship is as exact as possible (i.e. the gradient pulses are as short as possible). In practice, and particularly in clinical applications, this requirement is violated and the Fourier relationship is only approximate. Nevertheless, the $\mathbf{q}$-space approximation underpins much of the modelling literature and we will make use of it here. In the remainder of this article, the Fourier variable $\mathbf{q}$ should be interpreted as referring to $\mathbf{q}$, defined in \eq{eq:qdef} above.

\subsubsection{Solving the equations}
\label{solvevecdiff}
We have seen that the vector differential equations of interest to us here can be written in the form of \eq{eq:vecdiff}. First order linear vector differential equations can usually be solved and this section outlines a useful approach, adapted from \cite{jacobs2010}. We start by noting that the 1D case is quite trivial. In this case we have a scalar equation with a single constant coefficient
\beq
    \pdd{m}{t}=Am
    \label{eq:scalarform}
\eeq
where $m=m(t)$ and B is constant. This is easily solved by, for example, separation of variables or the Laplace transform (see, e.g. \cite{vvdenskii}). This equation has solution
\beq
    m(t)=e^{At}m(0)
    \label{expform}
\eeq
where $m(0)$ is a constant determined by initial conditions. For higher dimensional systems with the form of \eq{eq:vecdiff} the situation is complicated by coupling terms between different elements. However, if we consider a system 
\beq
    \pdd{\mathbf{n}}{t}= D\mathbf{n}
    \label{eq:diagvecdiff}
\eeq
where
\beq
    D=\left(\begin{array}{cccc}
                \lambda_1 & 0 & \cdots & 0\\
                0 & \lambda_2 & \cdots & 0\\
                \vdots & \vdots & \ddots & \vdots\\
                0 & 0 & \cdots & \lambda_k\\
            \end{array}\right)
    \label{eq:diagmat}
\eeq
we can immediately see that this is a set of uncoupled equations, i.e. 
\beq
    \pdd{n_k}{t} = \lambda_kn_k.
\eeq
Each has the form of \eq{eq:scalarform}. This system therefore has solution
\beq
    \mathbf{n}(t)= \left(\begin{array}{cccc}
                        e^{\lambda_1t} & 0 & \cdots & 0\\
                        0 & e^{\lambda_2t} & \cdots & 0\\
                        \vdots & \vdots & \ddots & \vdots\\
                        0 & 0 & \cdots & e^{\lambda_kt}\\
    \end{array}\right)\mathbf{n}(0) = e^{Dt}\mathbf{n}(0).
    \label{eq:expdiag}
\eeq
We can relate this back to the general, coupled case by requiring that
\beq
    \mathbf{n}=U\mathbf{m}
\eeq
where $U$ is a unitary transform that decouples the resulting system of equations. Such a matrix is guaranteed to exist if $A$ is square and normal (i.e. $A^\dagger A= AA^\dagger $, where $A^\dagger $ is the Hermitian conjugate of $B$). Since $U$ is unitary, the solution of \eq{eq:vecdiff} is therefore
\beq
    \mathbf{m}(t)=U^\dagger e^{Dt}U\mathbf{m}(0).
    \label{eq:vecdiffsoln}
\eeq
This also defines the \emph{matrix exponential}
\beq
    e^{At}=U^\dagger e^{Dt}U.
    \label{eq:matexp}
\eeq
In practice, $U$ and $D$ are obtained from an eigen decomposition of $A$. $D$ is the diagonal matrix of eigenvalues, $U$ is the matrix of Eigenvectors of $B$. As mentioned in the previous section, if $B$ is composed of two or more component matrices (as in \eq{vecdiff_combo}), where one component is a diagonal matrix it will make an additive contribution to the Eigenvalue matrix $D$. This propagates into the exponent, and hence leads to a multiplicative weighting in the solution.

As a final remark, we note that in order to obtain an equation of the desired form it was necessary to take a spatial Fourier transform. Ordinarily, we would expect to have to invert this transform in order to obtain a solution in real space. In the context of the current work, however, the spatial Fourier transform corresponds to a $q$-space approximation of the measured signal \cite{callaghan}. Since this is a more convenient representation of the signal for data analysis, and given that the use of the $q$-space approximation is widespread in the diffusion MRI literature, we will not in general perform the inverse Fourier transform and leave the obtained solutions in $q$-space. 

The general scheme for solving vector differential equations we will use it therefore
\begin{enumerate}
    \item{Write the equation in the form of \eq{eq:vecdiff}.}
    \item{Diagonalise the matrix to give $U$ and $D$.}
    \item{Construct the matrix exponential and solution.}
\end{enumerate}

\section{The continuity equation}
Conservation laws and conserved quantities are familiar concepts to every physicist. The idea that the total amount of a particular quantity is conserved (i.e. the total amount of it does not change overall) in a particular system is central to almost all of physics, and to almost all theoretical and mathematical approaches to a huge variety of problems. The idea of continuity is an extension of conservation to include more local effects and prevent certain, unphysical behaviours that are not precluded by conservation alone. 
As a very general statement of the concept, we might formalise the conservation of a quantity $p$ in some domain $\Omega$ as
\beq
    \int_\Omega p d\Omega = C
    \label{eq:cons}
\eeq
where $C$ is a constant. Simply, the total amount of $p$ in $\Omega$ is constant. This statement precludes the (net) creation or destruction of $p$ in $\Omega$ but says nothing about how it is distributed or how it moves around. By itself, \eq{eq:cons} does nothing to stop individual pieces of $p$ ``teleporting'' from one place to another. I.e. discontinuous jumps are not precluded by \eq{eq:cons}.

Whilst conservation is very general and useful, in many situations it is helpful to preclude jumps in $p$.  The presence of jumps means that derivatives of $p$ are not defined, and hence approaches to studying its dynamics that rely on differential equations, such as the Bloch-Torrey equation, will fail. For these approaches to be valid, it is necessary to impose a condition that requires $p$ to vary smoothly in space and time. Continuity is just this notion -- it is a tighter form of the conservation principle which accounts for the motion of a quantity as well as the total amount. Notice that the assumption of smooth variation in quantities is the same as that made implicitly in the Bloch-Torrey equation. 

Continuity is a widespread approach in physics. The first usage was probably by Euler who used them in his studies of fluids as early as 1757 \cite{continuity}. This section describes how continuity is formalised for both scalar- and vector-valued quantities.

\subsection{Continuity of a scalar quantity}
We can think about this by considering the amount of a quantity in a particular control volume $V$. Let $\rho(\mathbf{r})$ be the density of some scalar quantity at a point $\mathbf{r}$ in space. The total amount of $\rho$ in an arbitrary volume $V$ at time $t$ is then
\beq
    p(t)=\iiint_V\rho(\mathbf{r},t) dV.
    \label{eq:q}
\eeq
Unlike the conservation law in \eq{eq:cons}, this quantity is not guaranteed to be constant. In fact, its rate of change can be readily defined as
\beq
    \dd{p(t)}{t}= \d{t}\iiint_V\rho(\mathbf{r},t) dV=\iiint_V\pdd{\rho(\mathbf{r},t)}{t} dV.
    \label{eq:change1}
\eeq

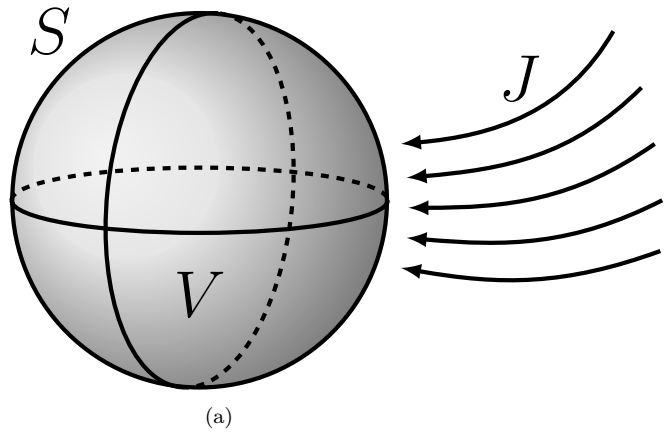
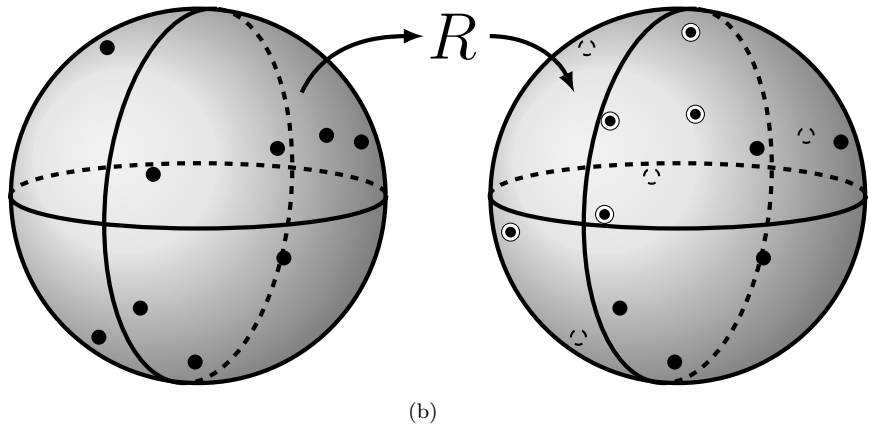
\begin{figure}
    \centering
    \subfloat[]{%
        \centering
        \begin{minipage}{0.45\textwidth}
            \begin{tikzpicture} 

            \def\R{2.5} 
            \def\angEl{10} 
            \pgfpathcircle{\pgfpoint{0cm}{0cm}}{28.3465*\R}						
            \pgfshadepath{ballshading}{20}
            \DrawLatitudeCircle[\R]{0.0}
            \DrawLongitudeCircle[\R]{60.0}

            \node at (-0.8*\R,0.9*\R){\Huge $S$};
            \node at (0.0,-0.5*\R){\Huge $V$};
            \node at (1.7*\R,0.65*\R){\Huge $J$};

            \draw[->,ultra thick] (2.2*\R,0.9*\R) to[out=240,in=5] (1.07*\R,0.3*\R);
            \draw[->,ultra thick] (2.35*\R,0.6*\R) to[out=225,in=5] (1.1*\R,0.12*\R);
            \draw[->,ultra thick] (2.42*\R,0.3*\R) to[out=215,in=0] (1.11*\R,-0.1);
            \draw[->,ultra thick] (2.46*\R,0.0*\R) to[out=207,in=355] (1.1*\R,-0.2*\R);
            \draw[->,ultra thick] (2.45*\R,-0.27*\R) to[out=200,in=350] (1.07*\R,-0.36*\R);

            \end{tikzpicture}
        \end{minipage}
    }
    \vspace{1in}
    \subfloat[]{
        \centering
        \begin{minipage}{0.45\textwidth}
            \begin{tikzpicture} %

            \def\r{0.1} 
            \def\R{2.5} 
            \def\angEl{10} 
            \pgfpathcircle{\pgfpoint{0cm}{0cm}}{28.3465*\R}						
            
            \pgfshadepath{ballshading}{20}
            \DrawLatitudeCircle[\R]{0.0}
            \DrawLongitudeCircle[\R]{60.0}


            \fill[black](0.86774*\R,0.2867*\R) circle(\r);
            \fill[black](0.45561*\R,-0.33035*\R) circle(\r);
            \fill[black](0.68268*\R,0.32205*\R) circle(\r);
            \fill[black](-0.48392*\R,0.78884*\R) circle(\r);
            \fill[black](-0.23911*\R,0.11454*\R) circle(\r);
            \fill[black](-0.52831*\R,-0.75221*\R) circle(\r);
            \fill[black](-0.01641*\R,-0.88435*\R) circle(\r);
            \fill[black](0.42145*\R,0.25241*\R) circle(\r);
            \fill[black](-0.30765*\R,-0.59736*\R) circle(\r);

            \draw[->,ultra thick] (0.55*\R, 0.55*\R) to[out=65,in=180] (1.2*\R,0.85*\R);

            \end{tikzpicture}
        \end{minipage}
        
        \begin{minipage}{0.45\textwidth}
            \begin{tikzpicture} %

            \def\R{2.5} 
            \def\r{0.1} 
            \def\angEl{10} 
            \pgfpathcircle{\pgfpoint{0cm}{0cm}}{28.3465*\R}						
            
            \pgfshadepath{ballshading}{20}
            \DrawLatitudeCircle[\R]{0.0}
            \DrawLongitudeCircle[\R]{60.0}

            \node at (-1.2*\R,0.85*\R){\Huge $R$};

            \fill[black](0.86774*\R,0.2867*\R) circle(\r);
            \fill[black](0.45561*\R,-0.33035*\R) circle(\r);
            \fill[black](-0.01641*\R,-0.88435*\R) circle(\r);
            \fill[black](0.42145*\R,0.25241*\R) circle(\r);
            \fill[black](-0.30765*\R,-0.59736*\R) circle(\r);
           
            \fill[black](-0.88911*\R,-0.19294*\R) circle(\r);
            \fill[black](0.06927*\R,0.87030*\R) circle(\r);
            \fill[black](-0.35960*\R,0.39836*\R) circle(\r);
            \fill[black](0.09371*\R,0.43452*\R) circle(\r);
            \fill[black](-0.38927*\R,-0.09885*\R) circle(\r);
            
            \draw[ultra thick,white](-0.88911*\R,-0.19294*\R) circle(\r);
            \draw[ultra thick,white](0.06927*\R,0.87030*\R) circle(\r);
            \draw[ultra thick,white](-0.35960*\R,0.39836*\R) circle(\r);
            \draw[ultra thick,white](0.09371*\R,0.43452*\R) circle(\r);
            \draw[ultra thick,white](-0.38927*\R,-0.09885*\R) circle(\r);

            \draw[thin,black](-0.88911*\R,-0.19294*\R) circle(1.2*\r);
            \draw[thin,black](0.06927*\R,0.87030*\R) circle(1.2*\r);
            \draw[thin,black](-0.35960*\R,0.39836*\R) circle(1.2*\r);
            \draw[thin,black](0.09371*\R,0.43452*\R) circle(1.2*\r);
            \draw[thin,black](-0.38927*\R,-0.09885*\R) circle(1.2*\r);

            \draw[thick,dashed](0.68268*\R,0.32205*\R) circle(\r);
            \draw[thick,dashed](-0.48392*\R,0.78884*\R) circle(\r);
            \draw[thick,dashed](-0.13911*\R,0.10454*\R) circle(\r);
            \draw[thick,dashed](-0.52831*\R,-0.75221*\R) circle(\r);

           \draw[->,ultra thick] (-1.0*\R, 0.85*\R) to[out=0,in=115] (-0.55*\R,0.55*\R);

            \end{tikzpicture}
        \end{minipage}
    }
    \caption{The two mechanisms by which the amount of a quantity in a region can change. (a) By moving into or out of the region via a flux process $J$, or (b) via intrinsic changes in the quantity itself via a source or sink process. Here, black particles decay away (dotted circles), and new particles emerge (ringed circles). Intrinsic change is described by the process $R$.}
    \label{continuity}
\end{figure}

This change can also be expressed in terms of the processes that cause it. The change in the total amount of a quantity in the volume $V$ is given by the net amount of the quantity entering and leaving the volume, and also by the intrinsic change in the amount of quantity present due to its creation and destruction. \Fig{continuity} illustrates these two processes -- a process transferring a quantity into (or out of) a region is known as a flux (a). A quantity may also change in the absence of movement (b) for example chemical reactions can occur or radioactive decay may transform on quantity into another.

The amount of $\rho$ entering or leaving the region can be expressed as a \emph{flux} -- an expression that describes how the quantity is moving around. Let $\mathbf{J}(\mathbf{r},t)$ be the net vectorial flux of $p$ in to/out of a point $\mathbf{r}$ and time $t$, and define the net intrinsic change in the quantity in the volume at time $t$ as $\Sigma(t)$. This means we can write the net change in $p(t)$ as
\beq
    \dd{p(t)}{t}= \oiint_S \mathbf{j}(\mathbf{r},t) dS + \Sigma(t)
    \label{eq:change2}
\eeq
where $S$ is the surface that bounds $V$. Equating \eq{eq:change1} and \eq{eq:change2} we have
\beq
    \iiint_V\pdd{\rho(\mathbf{r},t)}{t} dV= \oiint_S \mathbf{J}(\mathbf{r},t) dS + \Sigma(t).
    \label{eq:change3}
\eeq
With a small amount of manipulation we can put this in a more convenient form. Firstly we note apply the divergence theorem \cite{divthm} and rewrite the flux term as
\beq
\oiint_S  \mathbf{J}(\mathbf{r},t) dS = \iiint_V \div{\mathbf{J}(\mathbf{r},t)}dV
    \label{eq:divThm}
\eeq
and the source term can be written in terms of a local density
\beq
    \Sigma(t)= \iiint_V\sigma(\mathrm{r},t) dV.
\eeq
Substituting these into \eq{eq:change3} gives
\beq
\iiint_V \pdd{\rho(\mathbf{r},t)}{t} dV= \iiint_V \div{\mathbf{J}(\mathbf{r},t)}dV +  \iiint_V\sigma(\mathrm{r},t) dV.
    \label{eq:change4}
\eeq
Notice now that all three terms now contain volume integrals over $V$. Since $V$ is arbitrary, the only way in which \eq{eq:change4} can hold is if the integrands are equal, i.e.
\beq
    \pdd{\rho(\mathbf{r},t)}{t}=  \div{\mathbf{J}(\mathbf{r},t)} +  \sigma(\mathrm{r},t). 
    \label{eq:cons_scalar}
\eeq
This is the \emph{continuity equation}. It captures the changes in a conserved quantity in the presence of a flux, $\mathbf{J}$ and a source and sink process $\sigma$. The continuity equation can be applied to physical quantities such as mass, energy, or charge and also to other, less physical quantities like probability. This relation is exact and furthermore does not require that the spatial distribution of the quantity be continuous -- it is as useful for collections of particles as it is for continua.

The crucial concept for our current purposes, though, is the picture of a continuum or ensemble of particles, each ``tagged'' with some quantity of interest. The total amount of the tagged quantity at a particular location is controlled by two processes: a transport process defined by the flux, and a source/sink process which describes intrinsic changes in the tagged quantity. As yet, we have made no assumptions about the nature of this quantity other than the fact that it is a scalar. We could, for example, consider a physical quantity like mass or charge, or we could potentially consider another concept, like probability. All of these would amenable to analysis based on the continuity equation.

\subsection{Continuity of a vector quantity}
The above derivation is interesting, but in the context of MR physics it has a significant shortcoming: we have assumed that the quantity of interest can be described by a scalar. In MR physics we are interested in particles tagged with a vector quantity: Magnetisation. We might therefore ask if it is possible to adapt the above argument to vector-tagged particles. The answer, as it turns out, is yes. Conservation of vector quantities is a well-established idea: momentum is a conserved vector quantity, and considering the continuity of momentum in fluid dynamics leads directly to the Navier-Stokes equations \cite{fluids}. The same concept applies in continuum descriptions of non-fluids, and is also fundamental in the continuum mechanics of solids and gel-like media and elastography \cite{continuumMech}.

In fact, momentum is a more complex case than we need to consider, since there is a coupling between the conserved quantity motion. We can derive a continuity equation for a vector quantity that is not coupled to transport relatively straightforwardly. We will demonstrate this by considering each term in turn. 

The time derivative generalises trivially to the vector case: replacing $\rho$ by a vector quantity $\mathbf{M}$ yields a vector derivative. The same is true of the source and sink terms. We can generalise $\sigma(t)$ to describe the change of a vector quantity and define it as the density of some spatially extended quantity in the same way as the scalar case. The flux term, however, is a little more complicated.

The flux term defines the change in a quantity due to the motion of tagged particles. We derived the form of the flux by using the divergence theorem to relate a surface integral to a volume integral. It is not immediately clear how to define the flux of a vector quantity -- what sort of object describes this notion? This question is encountered in a different context in the Fluid Dynamics literature. The transport of a vector quantity means that there is a transport vector multiplying each element of the conserved quantity. This is well-expressed by the Dyadic product, giving the flux term the form
\beq
    \oiint_S \mathbf{JM} dS.
    \label{eq:flux-dyad}
\eeq
This is a higher rank expression than the vector quantities expressed in the divergence theorem, but as we have seen from section section \ref{tens-contr}, the divergence theorem generalises to higher rank objects as a Tensor Contraction. We can therefore re-write \eq{eq:flux-dyad} as
\beq
    \oiint_S \mathbf{JM}dS= \iiint_V \div{\mathbf{JM}}dV
\eeq
where the divergence operator sums over the flux index and the vector form of the continuity equation is therefore
\beq
	\pdd{\mathbf{M}}{t}+\div{\mathbf{JM}} = \mathbf{\sigma}(t)
    \label{eq:cont-vec}
\eeq

\begin{figure}
    \begin{center}
        \begin{pspicture}(-2,-1)(3,4)

            \psset{viewpoint=100 45 30,Decran=100}

            \pstVerb{
                /gro {
                    4 dict begin
                        /M defpoint3d
                        /a .5 def
                        /b 1 a 3 sqrt mul sub def
                        /k M norme3d a mul b add def
                        M k mulv3d
                    end
                } def}%

            \psSolid[object=grille, linestyle=dotted, linewidth=0.15pt, linecolor=lightgray, base=0 5 0 5, ngrid=10 10]%
            \psSolid[object=grille, linestyle=dotted, linewidth=0.15pt, linecolor=lightgray,base=-5 0 0 5, ngrid=10 10, RotY=90]%
            \psSolid[object=grille, linestyle=dotted, linewidth=0.15pt, linecolor=lightgray, base=-5 0 -5 0, ngrid=10 10,  RotY=90, RotZ=90]%

            \axesIIID[linewidth=1.5pt,
                arrowsize=5pt, 
                arrowinset=0,
                axisnames={\Delta x,\Delta y,\Delta z},
                labelsep=10pt]
            (0,0,0)(6,6,6)

    
            \psSolid[object=cube, a=5, linewidth=1pt, action=draw](2.5 2.5 2.5)

            \psSolid[object=cube, fillcolor=lightgray, a=0.5, action=draw*](2.25 2.25 2.25) 

            \psSolid[object=grille, base=0 5 0 5, ngrid=2 2, linestyle=dotted, linewidth=0.15pt, action=draw](0 0 5)
            \psSolid[object=grille, base=-5 0 0 5, ngrid=2 2, RotY=90, linestyle=dotted, linewidth=0.15pt, action=draw](5 0 0)%
            \psSolid[object=grille, base=-5 0 -5 0, ngrid=2 2,  RotY=90, RotZ=90, linestyle=dotted, linewidth=0.15pt, action=draw](0 5 0)%


            \psSolid[object=vecteur, args=0 -4.75 -0.9, linewidth=1.5pt](2.75 7.5 3.25) 

            \rput[br]{*0}(6.1,1.5){$\Delta\mathbf{M}$}

        \end{pspicture}
        \caption{A continuum of sub-volumes, each of which is labelled with a vector quantity. The net vector for the extended region is a sum over all sub-volumes. Such a continuum could represent a fluid, solid, or any material that is well-described by continuous variables.}
        \label{cont-vec}
    \end{center}
\end{figure}
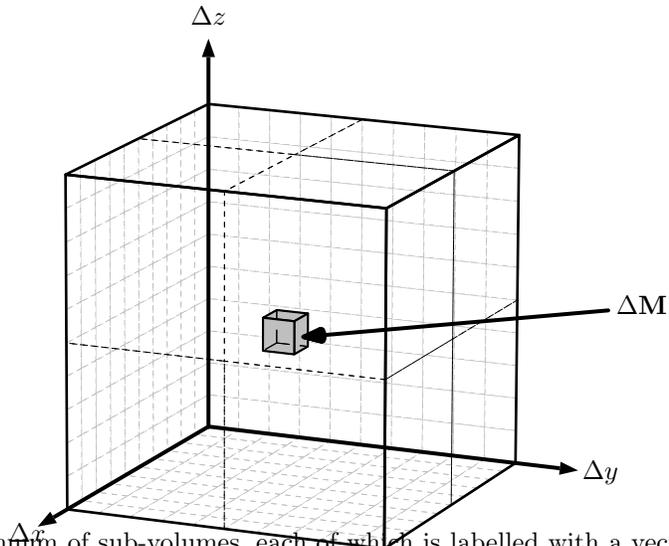

Notice also how the the idea of particles being tagged with a particular quantity of interest is preserved. Just as \eq{eq:cons_scalar} can be used to describe ensembles or continua tagged with a scalar measure such as mass or charge, \eq{eq:cont-vec} describes ensembles or continua or particles tagged with a vector quantity which changes from place to place according to local density, movement, and an intrinsic evolution process. This is model of a continuum in which locations in space correspond to infinitesimal sub-volumes, each of which is labelled with a vector quantity (see \fig{cont-vec})

\section{Continuity of magnetisation: the Bloch-Torrey equation}
We will now show that the the Bloch and Bloch-Torrey equations, which are fundamental to NMR and MRI experiments may be derived from the continuity of a vector quantity, and are in fact both particular cases of the continuity of magnetisation in the presence of a strong applied field.

\subsection{The Bloch terms as sources and sinks}
\label{blochterms}
The source and sink terms of the continuity equation describe the evolution of the quantity of interest that occurs independently of transport in the medium -- for example radioactive decay of a particular isotope causes a change in its local concentration that is independent of whether particles are moving or not. For scalar quantities the range of possible effects is quite limited: a scalar quantity can either increase or decrease, or in some cases change sign but no other changes are possible -- hence the origin of the terminology. Source terms lead to a local increase, sink terms to a decrease. 

For vector quantities, however, there are more possibilities: a vector may change not only magnitude but also orientation. In this case the notion of source and sink terms capture a wider class of processes. In the present context, they capture the interaction between a spin's magnetisation and the local magnetic field. Exactly the role of the terms of the Bloch equation. We may therefore consider the Bloch equation as a set of source and sink terms for our vector continuity equation.

We can therefore write \eq{eq:cont-vec} as
\beq
    \dd{\mathbf{M}}{t}= \gamma \mathbf{M}\times\mathbf{B} - \frac{M_x\mathbf{i} + M_y\mathbf{J}}{T_2} - \frac{M_z-M_0}{T_1}\mathbf{k} - \div{J(\mathbf{M})},
    \label{eq:btgen_old}
\eeq
i.e. a vector continuity equation with sources and sinks from the Bloch equation. We can immediately see the similarity between \eq{eq:btgen_old} and the traditional form of the Bloch-Torrey equation. \Eq{eq:btgen_old} describes the change in magnetisation of a continuous quantity in an applied magnetic field with an arbitrary transport process given by $J(\mathbf{M})$. This is, of course, the same approach used by Torrey in deriving his equation in the first place. The difference being that Torrey does not start from the general form with an unspecified $J(\mathbf{M})$ but instead proceeds directly to the drift-diffusion case \cite{torrey}. \eq{eq:btgen_old} contains the same physics, just with one less assumption.

We will see later that different choices of flux describe different NMR contrast mechanisms. Before we consider the full version of this equation and various choices for the flux, it it worth considering the Bloch terms by themselves, and in particular the notation we employ to describe them. The will simplify our discussion later, and will also help better illuminate the behaviour of the Bloch terms in a modern theoretical treatment. 

We begin by noting that \eq{eq:bloch} employs unit vector notation. This is of course perfectly valid, but obscures the essentially simple nature of the equation. First we consider the relaxation terms. The $T_2$ terms are the product of an element of the magnetisation with a time-like constant. The $T_1$ term is similar, but in addition to the product of a magnetisation element with a time-like constant, there is also an additive contribution with a constant (the $M_0$ term). The result of these products and their unit-vector sum is necessarily a vector in order to make sense when equated with the vectorial time derivative on the left hand side. We might therefore ask: what sort of product is expressed on the right hand side.

This form is analogous to the discussion of affine transforms in \secn{homog}. We have relaxation processes that are both additive and multiplicative. As such, we can employ an augmented matrix approach as re-write these terms as 
\beq
	\frac{M_x\mathbf{i} + M_y\mathbf{j}}{T_2} - \frac{M_z-M_0}{T_1}\mathbf{k} 
																\to \left(\begin{array}{cccc}
																		-\frac{1}{T_2} & 0 & 0 & 0 \\
																		0 & -\frac{1}{T_2} & 0 & 0 \\
																		0 & 0 & -\frac{1}{T_1} & \frac{M_0}{T_1} \\
																		0 & 0 & 0 & 1\\
																	\end{array}\right)
																		\left(\begin{array}{c}
																			M_x\\M_y\\M_z\\1\\
																		\end{array}\right).
\eeq
The Bloch terms also contain a precessional term, which takes the form of a cross product with the applied field. We can write this in a more compact form by noting that a cross product can be written as a matrix multiplication. Specifically,
\beq
	\mathbf{a}\times\mathbf{b} = \left(\begin{array}{ccc}
									0 & -a_z & a_y \\
									a_z & 0 & -a_x \\
									-a_y & a_x & 0 \\
								 \end{array}\right)
									\left(\begin{array}{c}
										b_x\\ b_y \\ b_z\\
									\end{array}\right)
\eeq
which can be verified by expanding the terms on both sides.\footnote{In fact, this relationship is deeper than it first appears. It provides a mapping between vector multiplication and the group $SO(3)$. It connects directly with Lie groups and Lie algebras, but this is beyond the scope of the current work.}. 
This is a skew-symmetric matrix -- it has anti-symmetric off-diagonal elements. Interestingly, it is non-zero in elements in which the relaxation matrix is zero an vice versa. If we choose, we can embed this term in the augmented matrix, thus giving an complete description of the Bloch equation as a single matrix multiplication.
\beq
	\left(\begin{array}{c}\dd{M_x}{t}\\\dd{M_y}{t}\\\dd{M_z}{t}\\1\end{array} \right)
	=\left(\begin{array}{cccc} -\frac{1}{T_2} & -\gamma B_z & \gamma B_y & 0\\
				 	\gamma B_z & -\frac{1}{T_2} & -\gamma B_x & 0\\
					-\gamma B_y & \gamma B_z & -\frac{1}{T_1} & \frac{M_0}{T_1} \\
					0 & 0 & 0 & 1
		\end{array}\right)
				\left(\begin{array}{c}M_x\\M_y\\M_z\\1\end{array} \right)= R\mathbf{M},
	\label{eq:matbloch}
\eeq
which may be simplified in the absence of applied field gradients since we usually take the direction of the applied field as the definition of the lab frame's $z$-axis. In this case we have,
\beq
	\left(\begin{array}{c}\dd{M_x}{t}\\\dd{M_y}{t}\\\dd{M_z}{t}\\1\end{array} \right)
	=\left(\begin{array}{cccc} -\frac{1}{T_2} & -\gamma B_z & 0 & 0\\
				 	\gamma B_z & -\frac{1}{T_2} & 0 & 0\\
					0 & \gamma B_z & -\frac{1}{T_1} & \frac{M_0}{T_1} \\
					0 & 0 & 0 & 1
		\end{array}\right)
				\left(\begin{array}{c}M_x\\M_y\\M_z\\1\end{array} \right)= R\mathbf{M},
	\label{eq:matbloch2}
\eeq
where the applied field is now parallel to the $z$-axis.

This expresses the Bloch equation in a more modern notation at the expense of adding a trivial fourth equation which we can safely ignore. This notation immediately makes two facts apparent: firstly, the time increment can be expressed as a projection of a point in magnetisation space into a tangent ``magnetisation increment space''. Secondly, the projection itself is an affine transformation: there is scaling operation (the diagonal relaxation terms), the terms from the cross product unsurprisingly manifest themselves as a rotation, and a translation is applied parallel to the z-axis (the fourth column, containing the $M_0$ term). Since the repeated application of scaling operations will exponentially decrease the length of the original vector, we can see immediately that in the long time limit (i.e. $t \gg T_1, T_2$) the translation term will dominate, means that it defines an equilibrium state on which the system rapidly converges. We can also immediately see that the equilibrium magnetisation vector is parallel to the $z$-axis, since the translation terms have no $x$ or $y$ components -- all this before we attempt to actually solve the equation!

Before we leave this section it is worth making a remark about the rotating frame of reference. We have seen that the Bloch terms include a rotational, cross product term which describes the precession of the magnetisation vector about the applied field. We have seen that this can be written in matrix form and added to the Bloch matrix. This leads to a multiplicative factor in the solution which rotate the system about the $z$ axis by an angle of equal to $\gamma B_z t$. Transforming to a co-rotating frame of reference leads to this cross product term being included into the time derivative, and hence cancelling these terms from the Bloch matrix \cite{rotframe}. In the current context this leads only to a change of representation in the Bloch matrix and hence its Eigenvalues and exponent - the methodology we use to solve the equations is unchanged. In the remainder of this work, we will denote the Bloch terms as the matrix multiplication $R\mathbf{M}$, noting that a change of coordinates will affect the elements of $R$ but not the form of the multiplication itself.

\subsection{The flux terms: transport processes}
By substitution in \eq{eq:btgen_old} in a more compact form,
\beq
    \pdd{\mathbf{M}}{t} = R\mathbf{M} -J(\mathbf{M}).
    \label{eq:btgen}
\eeq
$R$ contains the Bloch terms defined in \eq{eq:matbloch2}, and $J(\mathbf{M})$ is a general flux term describing a transport process of interest. \Eq{eq:btgen} is \eq{eq:btgen_old} written in a more compact form, and is a generalised form or the Bloch-Torrey equation with a generalised flux term.

Different choices of flux lead to different solution of \eq{eq:btgen}, and in many cases to different specific MR imaging techniques. We can immediately see that setting $J(\mathbf{M})=0$, the case assuming stationary spins, recovers the Bloch equation. The Bloch-Torrey equation is recovered by assuming Fick's law, $J(\mathbf{M})= D\grad{\mathbf{M}}$, which assumes that diffusion is Markovian, and transport from regions of high concentration to low, proportional to the linear gradient. Here $D$ is the diffusion tensor, which defines the orientational dependence of the diffusive process, and is the object of interest in Diffusion Tensor Imaging (DTI). 

We will see in the next section that by exploring different choices of flux this equation provides a single framework that unifies much of the work in diffusion MRI over the last decade or so, and also includes other modalities which are sensitive to net spin flux. In the remainder of this paper we will outline how these different techniques are related to each other, and then outline some potential new research directions that are suggested by the idea of focusing on spin transport. Different choices of flux and the resulting model are summarised in table-\ref{fluxes}.

The $\mathbf{q}$-space formalism casts all our equations in terms of displacements. Hence we assume that all spins start with displacement zero and then move according to the specified flux. We are also implicitly assuming that there are not explicit barriers to the flux. Although this assumption is widespread, it is not common to all models of Diffusion MRI. In particular the models discussed in Section \ref{with_bounds} make strong assumptions about geometric boundary conditions, but we will see that this simple free-space assumption is compatible with much of the modelling literature.

\section{Flux choices and other imaging modalities}
This section reviews the diffusion MRI modelling literature in the context of the theoretic framework developed above. We will show how many existing techniques stem from choices of flux terms of generalised Bloch-Torrey equations and demonstrates that the framework acts as a unifying principle across what can otherwise appear a bewildering maze of different models. We will also see that the common assumption that the phase information is not important is only valid in particular special cases.

\begin{table}[ht]
    \caption{Flux term choices and corresponding diffusion models}
    \label{fluxes}
    \begin{center}
        \begin{tabular}{ c | c | c }
            \hline
            Flux & Form & Technique \\
            \hline
            0   &  (no diffusion terms) & $T_1$ and $T_2$ weighting \\
            $\mathbf{uM}$ & $e^{i\mathbf{u}\cdot\mathbf{q}}$ & \small{Velocity-weighted phase contrast}\\
            $-D\grad{\mathbf{M}}$ & $e^{-\mathbf{q}D\mathbf{q}t}$ &  \small{Diffusion tensor imaging}\\
            $-D\frac{\partial^\beta}{\partial |x|^\beta}\mathbf{M}$ & $e^{-D|q|^\beta t}$ & \small{Stretched exponential}\\
            (transfer kernel) & \small{$e^{\iint T(\mathbf{q})\hat{\theta}(\mathbf{q},t)\cos\theta'd\theta'd\phi}$} & \small{Spherical deconvolution}\\
            \tiny{$D\grad{\mathbf{M}} -\frac{1}{3}D\grad{K}\div{D}\grad{\mathbf{M}}t$} & \tiny{$e^{-\mathbf{q}D\mathbf{q}t}e^{-\frac{1}{6}\mathbf{q}D\mathbf{q}K\mathbf{q}D\mathbf{q}t^2}$} & \small{Diffusion Kurtosis Imaging}\\
            \hline
        \end{tabular}
    \end{center}
\end{table}

\subsection{$T_1$- and $T_2$-weighted imaging}
\label{solvebloch}
This section solves the vector form of the Bloch equation. We do this partly as an illustration of the method illustrated above, and partly as a foundation for the remaining discussion -- the solution of the Bloch equation forms part of the solutions for all other models discussed subsequently and as such is of interest to each of them.

In the vector notation we developed previously, the Bloch equation is written as
\beq
    \pdd{\mathbf{M}}{t} = R\mathbf{M}
\eeq
where
\beq
     R= \left(\begin{array}{cccc}
            -\frac{1}{T_2} & 0 & 0 & 0 \\
			0 & -\frac{1}{T_2} & 0 & 0 \\
			0 & 0 & -\frac{1}{T_1} & \frac{M_0}{T_1} \\
			0 & 0 & 0 & 1\\
	\end{array}\right)
\eeq
in the rotating frame.

We know that the solution of the Bloch equation is given by \eq{eq:vecdiffsoln}, so our task reduces to obtaining the matrix exponential of $R$, as defined in \eq{eq:matexp}. Diagonalising this matrix gives us the set of Eigenvectors
\beq
    R_V=\left(\begin{array}{cccc}
            0 & 0 & 1 & 0\\
            0 & 1 & 0 & 0\\
           -M_0 & 0 & 0 & 0\\
            1 & 0 & 0 & 0\\
    \end{array}\right)
\eeq
which has the inverse
\beq
    R_V^{-1}= \left(\begin{array}{cccc}
            0 & 0 & 0 & 1\\
            0 & 1 & 0 & 0\\
            1 & 0 & 0 & 0\\
            0 & 0 & 1 & M_0\\
    \end{array}\right)
\eeq    
and eigenvalues
\beq
    R_D=\left(\begin{array}{cccc}
        0 & 0 & 0 & 0\\
        0 & -\frac{1}{T_2} & 0 & 0\\
        0 & 0 & -\frac{1}{T_2} & 0\\
        0 & 0 & 0 &  -\frac{1}{T_1}\\
   \end{array}\right).
\eeq
The eigenvalues of the matrix exponential $\exp{R_Dt}$ is therefore
\beq
    e^{R_Dt} =\left(\begin{array}{cccc}
            1 & 0 & 0 & 0\\
            0 & e^{-\frac{t}{T_2}} & 0 & 0\\
            0 & 0 & e^{-\frac{t}{T_2}} & 0\\
            0 & 0 & 0 &  e^{-\frac{t}{T_1}}\\
   \end{array}\right)
\eeq
and the solution of the Bloch equation is
\beq
    \mathbf{M}(t) = R_Ve^{R_Dt}R_V^{-1}\mathbf{M}_0 = \left(\begin{array}{cccc}
                e^{-\frac{t}{T_2}} & 0 & 0 & 0\\
                0 & e^{-\frac{t}{T_2}} & 0 & 0\\
                0 & 0 & M_0(1-e^{-\frac{t}{T_1}}) & 0\\
                0 & 0 & 0 & 1\\
    \end{array}\right)
\eeq
which is the familiar expression describing $T_1$ and $T_2$ decay written is vector form.
 
This expression has been derived in the rotating frame, and thus describes the length of the magnetisation vector only. This is nevertheless sufficient to describe attenuation at echo time, since when $t=TE$ the rotating frame is consonant with the lab frame \cite{callaghan}. Adding the off-diagonal $B_z$ terms to the $R$ matrix simply leads to a multiplicative rotational term in the solution describing the change in coordinates, which we omit here. 

\subsection{Diffusion tensor imaging}
The next simplest case is Diffusion Tensor Imaging \cite{basser1994}. DTI is directly related to the canonical form of the Bloch-Torrey equation and the classical theory of diffusion. We first note that diffusion is defined as a flux of particles from regions of high concentration to low, and is characterised by Fick's Law
\beq
    J=-D\grad{C}
    \label{eq:fick1}
\eeq
where $J$ is the flux, $D$ is a rate-defining constant and $C$ is the local concentration of diffusing particles. We can relate this to the case where we have particles tagged with a magnetisation vector by defining the local magnetisation as the total dipole moment per unit volume. Assuming local equilibrium, the strength of the local magnetisation is then proportional to the local particle concentration and we can write a vectorial version of \eq{eq:fick1}
\beq
    J(\mathbf{M})=-D\grad{\mathbf{M}}.
    \label{eq:fick2}
\eeq
Notice that now the gradient operator is acting on a vector, and hence has a slightly different form to that in \eq{eq:fick1}. Generalising this operator is straightforward, and is defined as the outer product of the gradient and the magnetisation vector (this form is also used -implicitly- by Torrey \cite{torrey}, although it is not named as such). There is a choice to  be made regarding the form of $D$. In the case most usually employed in diffusion MRI, $D$ is a tensor which defines an orientational dependence on the local diffusion.

Substituting \eq{eq:fick2} into \eq{eq:btgen} gives the familiar form of the Bloch-Torrey equation. We can solve this using the method from \secn{solvevecdiff}. The Bloch terms from the $R$ matrix are the same as in \secn{solvebloch}, applying the Fourier Transform from \eq{eq:ft_q} to the flux term yields
\beq
FT[\div{J(\mathbf{M})}] = -FT[\div{D\grad{\mathbf{M}}}] = -i\mathbf{q}Di\mathbf{q}\mathbf{m}=\mathbf{q}D\mathbf{q}\mathbf{m}
\eeq
which in turn yields a diffusion-weighted attenuation term in the solution with the form
\beq
    e^{-\mathbf{q}D\mathbf{q}t},
\eeq
the usual, mono-exponential decay prediction. The complete solution is therefore
\beq
    \mathbf{m}(t)=\mathbf{m}_0e^{Rt}e^{-\mathbf{q}D\mathbf{q}t}.
    \label{eq:dti}
\eeq
It is common not to write the decay term explicitly and simply call the Bloch equation pre-factor $S_0$ and the measured signal $S$. $S$ is related to $\mathbf{M}$ via an integral over the local spin phase distribution, which is also implicit in the $q$-space approximation \cite{signal_to_magnitude}.

Determining the elements of $D$ requires several measurements of this attenuation term. This is usually performed by formulating and inverting a linear system, the structure of which is derived from this solution (see, e.g. \cite{lebihanDTI}).

For our current purposes, it is enough to see that DTI is the result of the assumption of a Fickian flux and a tensorial diffusivity, and to note that the attenuation factor is entirely real-valued, with no imaginary terms.

\subsection{Velocity-weighted phase contrast}
Another straightforward case we can consider is where spin transport is completely coherent, and has a well-defined, finite mean direction. Under these circumstances, spins can be thought of like a tracer dye or isotope: test particles used as a probe of local flow. This is equivalent to velocity-weighted imaging. Although a drift term was originally considered by Torrey \cite{torrey}, coherent flow effects are not traditionally considered alongside diffusion-weighted methods. This contrast mechanism, however, sits easily within the spin-transport framework we have developed and provides a useful contribution.

Again, we start with assuming a form for the flux term. This time we assume there is a local flow of spins which is described by a vector $\mathbf{u}$, giving us the following form for the flux
\beq
    J(\mathbf{M})= \mathbf{uM}.
    \label{eq:velflux}
\eeq
Again, this term involves an outer product. It describes the transport of each element of the magnetisation vector by each component of the velocity vector. Substituting this flux into \eq{eq:btgen} gives
\beq
    \pdd{\mathbf{M}}{t}= R\mathbf{M} - \div{\mathbf{uM}}.
    \label{eq:velwtd}
\eeq
Once again, the Bloch terms manifest themselves as in \eq{eq:matbloch}. Taking the Fourier Transform, we obtain
\beq
    \pdd{\mathbf{m}}{t}= R\mathbf{m} -i\mathbf{q}\cdot\mathbf{um}
\eeq
and we find the solution
\beq
    \mathbf{m}(t)=\mathbf{m}_0e^{Rt}e^{-i\mathbf{q}\cdot\mathbf{u}t}.
\eeq
i.e. the solution of the Bloch equation with an additional attenuation proportional to the dot product of the applied gradient (via the $\mathbf{q}$ vector) and the local mean velocity. 

Note here that the exponent in the velocity attenuation is imaginary, which means it has the form of a rotation in the complex plane and thus imparting a linear phase change to the signal proportional to the local mean velocity.

\psset{yunit=4cm,xunit=1.45}
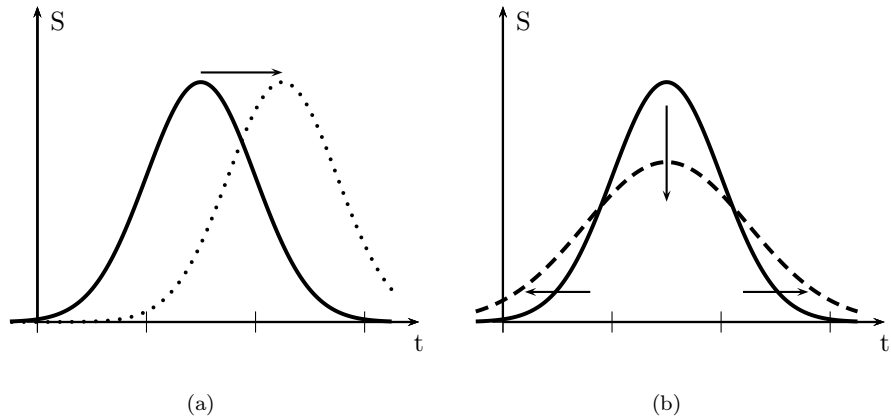
\begin{figure}[ht]
    \centering
        \subfloat[]{
            \centering
            \begin{pspicture}(-2,-0.2)(2,1.4)
                \psaxes[labels=none]{->}(-1.5,0)(2,1.05)
                \uput[-90](2,0){t}\uput[0](-1.5,1){S}
            

                \psGauss[linecolor=black, linewidth=1.5pt]{-1.75}{1.75}%
                \psGauss[linecolor=black, linestyle=dotted, mue=0.75, linewidth=1.5pt]{-1.75}{1.75}%

                \psline[linewidth=0.8pt]{->}(0,0.83)(0.75,0.83)
            \end{pspicture}
        }
        ~
        \subfloat[]{
            \centering
            \begin{pspicture}(-2,-0.2)(2,1.4)
                \psaxes[labels=none]{->}(-1.5,0)(2,1.05)
                \uput[-90](2,0){t}\uput[0](-1.5,1){S}
            
                \psGauss[linecolor=black, linewidth=1.5pt]{-1.75}{1.75}%
                \psGauss[sigma=0.75, linecolor=black, linestyle=dashed, linewidth=1.5pt]{-1.75}{1.75}

                \psline[linewidth=0.8pt]{->}(0,0.72)(0,0.4)
                \psline[linewidth=0.8pt]{->}(0.7,0.1)(1.3,0.1)
                \psline[linewidth=0.8pt]{->}(-0.7,0.1)(-1.3,0.1)

            \end{pspicture}
        }
        \caption{The difference between coherent (a) and incoherent (b) motion on the observed echo (arbitrary units on both axes). (a) Coherent motion leads to an increase in the mean, which changes the overall timing of the pulse and an effect on signal phase. (b) Incoherent motion leaves the pulse timing unchanged, but spreads it out leading to a reduction in pulse height, observed as a change in magnitude. These two effects are different degrees of freedom for the pulse, and are thus mathematically separable.}
        \label{advec}
\end{figure}

The combination of coherent flow and diffusion is equivalent to an advection-diffusion equation where spins both flow and diffuse but the two processes are essentially uncoupled. \Fig{advec} illustrates the effect on the signal of both coherent and incoherent motion.  Notice how the signal contrast predicted for Fickian diffusion (incoherent motion) is purely real, whereas the contrast predicted for coherent flow is purely imaginary. This suggests that contrast from the two mechanisms is in some sense orthogonal and could potentially be measured simultaneously and separated in an analysis of the full, complex MR measurements.

\subsection{Stretched-exponentials and space-fractional super-diffusion}
\label{sec:strexp}
The previous techniques all assumed a straightforward definition of the spin flux -- a coherent spatial shift leads to a coherent phase shift, equilibrating fluxes along gradients in concentration lead to conventional diffusion attenuation. It is possible, however, to make other choices which are indicative of more exotic spin transport processes. This section derives the signal attenuation due to a population of spins executing L\'evy walks. 

At the microscopic level, diffusing particles are modelled by assuming each one is undergoing Brownian motion. This assumes that the displacement of each particle in a given time interval is independent of the displacements in previous intervals -- the process is Markovian \cite{diffndynamicsref}. The microscopic dynamics can be shown to lead to Fick's law and the diffusion equation in a suitable limit. 

Diffusion from Brownian motion underpins the bulk of the diffusion MR literature, but there are other choices that we can make. An alternative model of microscopic diffusion-like dynamics is the L\'evy walk. Particles executing L\'evy walks make many short steps with the occasional much longer displacements. They are characterised by random walk-like processes with a power-law distribution of step lengths. I.e. step length $\ell$ follows
\beq
    p(\ell)\propto \frac{1}{\ell^{\beta}}.
    \label{eq:powlawsteps}
\eeq

L\'evy walks are good models of, for example, diffusion fractals in which diffusion is restricted across a wide range of length scales. Microscopic L\'evy dynamics can be shown to lead to fractional spatial derivatives in the continuum regime \cite{tomfractionalbook}. Just as a Brownian random walk leads to a Fickian flux, a similar calculation shows that a L\'evy walk leads to a flux with a fractional-order derivative. Fractional derivatives are derivatives of non-integer order. They interpolate between derivatives of integer order and are the subject of an extensive research literature, and have been extensively applied to transport theory. Whilst not providing a unique link between the microstructural details of the system, fractional approaches nonetheless provide access to systems in which diffusion-like transport it not well-described by conventional, regular diffusion. In this case, the order of the derivative is the same as the exponent on the power law.

As an example of L\'evy walks in diffusion MRI, we will consider the case where the flux term is described by a fractional derivative. For simplicity, we will consider only a one dimensional flux.

In this case, the flux is written as
\beq
    J(\mathbf{M})=-D\frac{\partial^{\beta}\mathbf{M}}{\partial |x|^\beta}
\eeq
where $D$ is a constant and $0<\beta<1$ is the order of the derivative (see Supplementary Material for a derivation of the fractional flux). The derivative operator has the Reisz form \cite{tomfractionalbook}, which describes a symmetric, bi-directional process. The governing equation for spins with this flux in 1D is therefore
\beq
    \pdd{\mathbf{M}}{t}= R\mathbf{M} + \pd{x}D\frac{\partial^{\beta}\mathbf{M}}{\partial |x|^\beta}
\eeq
which equals
\beq
\pdd{\mathbf{M}}{t}= R\mathbf{M} + D\frac{\partial^{\beta+1}\mathbf{M}}{\partial |x|^{\beta+1}}.
\eeq
The spatial FT gives
\beq
\pdd{\mathbf{m}}{t} = R\mathbf{m} - D|q|^{\beta+1}\mathbf{m}
\eeq
and the solution is
\beq
    \mathbf{m} = \mathbf{m}_0e^{Rt}e^{-D|q|^{\beta+1}t}.
	\label{strexp}
\eeq
This form of spin transport is known as super-diffusion, so-called because the mean squared-displacement of particles increases faster than linear with time (as would be the case with regular diffusion). The stretched exponential in this form is thus a model of super-diffusion.

Stretched exponentials have been employed by several authors \cite{bennett} \cite{silvia} \cite{hallBarrick}, although the form chosen varies. The form derived here has the stretching exponent on the $q$ terms only, although authors often apply the stretching exponent to both $q$ and $t$.

The model has been criticised on the grounds that the exponent is difficult to interpret physically, but the derivation here illustrates that it can be directly associated with the statistics of the underlying random walk: the fitted exponent is related to transport, not structure. 

Strictly speaking, this model describes (perhaps counter-intuitively) a model of diffusion in which the mean squared-displacement of spins increases \emph{faster} than linearly with time, which is difficult to justify as a model of diffusion in tissue. We would expect structure present in the environment, no matter how complex, to \emph{impede} spin motion and hence slow down the diffusion process. Superdiffusion, on the other hand, implies that some acts transport mechanism is driving spins apart faster than diffusion would otherwise allow.

\subsection{An alternative flux formulation: the transfer kernel}
So far we have always expressed the transfer of magnetisation into a given location using a flux (or, more precisely, the grad of a flux). In this section, however, we note that this is not the only way to express the net transfer of mass from one location to another. An alternative technique is to consider a process which models the net transfer between pairs of locations. A common approach to this is known as the Chapman-Kolmogorov transfer matrix \cite{codling}\cite{othmer}. This is simply a matrix defining the net transfer for point $\mathbf{x}$ to point $\mathbf{y}$ in a given time interval $dt$. For discrete systems, such as lattices, this is written as
\beq
\div{J(\mathbf{M}}) = \sum_j T_{ik}\mathbf{M}(\mathbf{x}_k,t)
    \label{ck-lattice}
\eeq
where $T_{ik}$ defines the net transfer between location $x_i$ to $x_k$. For continuous systems we can rewrite this as an integral over a control volume $V$, 
\beq
\div{J(\mathbf{M}}) = \iiint_V T(\mathbf{x},\mathbf{y}) \mathbf{P}(\mathbf{y},t)dV
    \label{ck-contm}
\eeq
where the matrix has now been replaced by a continuous kernel and $P(\mathbf{M},\mathbf{y},t)$ is a function which describes the distribution of $\mathbf{M}$ over the control volume. Note that $T(\mathbf{x},\mathbf{y})$ is a function of displacement only, i.e.
\beq
    T(\mathbf{x},\mathbf{y}) = T(\mathbf{x}-\mathbf{y})
    \label{disp-kernel}
\eeq
then the integral in \eq{ck-contm} has the form of a convolution between the transfer kernel and the distribution of magnetisation. In this case the Bloch-Torrey equation has the form
\beq
    \pdd{\mathbf{M}}{t} = R\mathbf{M} + \iiint_V T(\mathbf{x}-\mathbf{y})P(\mathbf{y},t)dV.
    \label{eq:BT-CK}
\eeq

This is a highly flexible description of the imaging equations. At this point we are free to make any choice we like for the transfer kernel $T$, and furthermore the convolution form is easy to work with in the $q$-space formalism that we have been using to solve the various equations. The Fourier transform of a convolution is a product \cite{vvdenskii}. 

\subsubsection{Spherical deconvolution}
The convolution formalism is also immediately suggestive of a widely-used method to resolve the so-called crossing fibre problem in DTI \cite{crossingfibres}. Spherical Deconvolution is a method developed by Tournier and co-workers \cite{SD}\cite{cSD}\cite{cSD_ISMRM2014}. It is motivated by a shortcoming of Diffusion Tensor Imaging -- specifically that it is only capable of resolving a single dominant diffusion direction per scan voxel. In many voxels in the brain, for example, there are complex, crossing-like configurations where more than one population of fibres is present \cite{crossings_freq}. Addressing this shortcoming led to the development of a suite of methods capable of resolving more complex fibre configurations, known collectively as High Angular Resolution Diffusion Imaging (HARDI). 

Spherical deconvolution models the diffusion-attenuated signal as convolution of a transfer kernel described by the diffusion signal response from a single population of parallel fibres convolved with a function describing the orientation distribution of fibres. The similarity with the above formalism is clear, however there are a number of additional assumptions necessary in order to express Spherical Deconvolution in the generalised Bloch-Torrey framework.

Firstly, HARDI methods are concerned primarily with directionality in the underlying diffusion measurements. As such the natural coordinates for the problem are spherical, rather than Cartesian. We also assume that the transfer kernel is axially symmetric and given by a single-population response function. This means that \eq{eq:BT-CK} may be re-written as
\beq
    \pdd{\mathbf{M}}{t} = R\mathbf{M} + \iiint_V T(\theta)P(r,\theta,\phi,t)\cos\theta' d\theta' d\phi r dr.
    \label{eq:BT-CK_sph}
\eeq
Furthermore, since we are only interested in the angular dependence of the signal, we assume that the $r$ dependence can be separated and integrated out. The usual procedure is to assume exponential decay in the radial part of the signal, integrate and assume this is a constant, i.e.
\beq
    \pdd{\mathbf{M}}{t} = R\mathbf{M} - \int r dr\iint T(\theta)\hat{\mathbf{\theta}}(r,\theta',\phi,t)\cos\theta' d\theta' d\phi
    \label{eq:BT-CK_sph2}
\eeq
i.e.
\beq
    \pdd{\mathbf{M}}{t} = R\mathbf{M} - C\iint T(\theta)\hat{\theta}(r,\theta',\phi,t)\cos\theta' d\theta' d\phi.
\eeq
where $\hat{\mathbf{\theta}}$ is the function describing the orientational distribution of fibre populations in the control volume. This is usually known as the Orientation Distribution Function (ODF). Typically this is considered only in the transverse plane. At this stage we can solve the equation to give a description of the diffusion-weighted signal due to the kernel which can be used to access the ODF which we are interested in estimating -- this solution makes a prediction which relates the ODF to the observed signal. Following the same procedure as in  previous sections we can see that the solution is
\beq
    \mathbf{m}(t)=\mathbf{m}_0e^{R\mathbf{m}}e^{-\iint T(\mathbf{q})\hat{\theta}(\mathbf{q},t)\cos\theta' d\theta' d\phi}
\eeq
The real challenge of all HARDI methods is, of course, how we solve the inverse problem: how do we get a particular ODF from a set of measurements. There is a rich literature on this topic, which is beyond the scope of this paper to review but in general we can distinguish between linear methods, which formulate a linear system in a similar way to in DTI and invert, and non-linear methods which perform a numerical optimisation procedure using the forward form and suitable cost function. Non-linear methods tend to give superior angular resolution in derived ODFs but are far more numerically intensive and require greater computational resource.

Other HARDI methods fit into the continuity equation picture in similar ways to spherical deconvolution. PAS-MRI \cite{pas} has been shown to be a deconvolution method via a related method, Maximum Entropy Deconvolution \cite{mesd}, and Q-ball is formulated as an integral transform which can also be seen as a particular choice of transfer matrix formalism (albeit a slightly less obvious one) \cite{qball1, qballproper}.

\subsubsection{Diffusion kurtosis imaging}
Diffusion Kurtosis Imaging is also motivated by shortcomings in DTI. Rather than focusing on the angular dependence of diffusion, DKI instead attempts to address the implicit assumption in DTI that the spin displacement distribution is Gaussian -- the Fickian flux leads directly to a Gaussian distribution for spin displacements. The solution we derived in \eq{eq:dti} is in fact the Fourier Transform of a 3D zero-mean Gaussian distribution. Gaussian distributions are observed in fluids but are known to be only approximations of spin displacement distributions in situations where spins are restricted by microstructure.

DKI seeks to provide a better approximation of non-Gaussian spin displacements by expanding the expression to the fourth-order moment of the distribution, Kurtosis (odd orders are assumed to be zero in the absence of net drift) \cite{dki}. This approach has the advantage that it makes no strong assumptions about what is causing the non-Gaussianity, but nonetheless quantifies it in a well-understood way. Indeed, non-zero excess Kurtosis is often described as being a signature of restriction \cite{kurtosis_as_restriction} but is has also been shown to be special case of fractional diffusion models \cite{carson_kurtosis}

Within the current framework, there are two ways to approach this. The first option is to express the Chapman Kolmogorov transfer kernel as an expansion in moments of the distribution, truncated at fourth order. 

The other method consists of constructing a flux term with an appropriate Kurtosis term. In this case the fourth order flux is as follows
\beq
    J(\mathbf{M}) = -D\grad{\mathbf{M}} -\frac{1}{3}D\grad{K}\div{D}\grad{\mathbf{M}}t
    \label{kurtflux}
\eeq
where $D$ is the second order diffusion tensor, and $K$ is the fourth order Kurtosis tensor which modifies the Gaussian flux process. The factor of $t$ is required to correct the dimensionality of the kurtosis term. Substituting this into \eq{eq:btgen} and solving yields the following solution
\beq
    \mathbf{m}(t)=\mathbf{m}_0e^{R\mathbf{M}}e^{-\mathbf{q}D\mathbf{q}t}e^{-\frac{1}{6}\mathbf{q}D\mathbf{q}K\mathbf{q}D\mathbf{q}t^2}
    \label{eq:kurtosis}
\eeq
which is the 3D form of the usual diffusion kurtosis solution. As with DTI, this can be used to construct a linear system over several measurements that can be inverted to find the elements of the diffusion and kurtosis tensors.

\subsection{Multiple continua}
\label{multi_cont}
So far we have considered models with only a single, well-mixed continuum is present in the sample. The linear transport literature, however, contains a large body of work in which we consider the effect of two or more continua in a model. This approach is well-suited to MR methods, since it assumes that measurements are made at relatively large spatial scales at which we can consider two or more compartments to be spatially well-mixed, without imposing specific spatial relationships between them. This assumption fits well with the idea that we are interested in measurements at the voxel scale, which contain spatially averaged microstructure which affects the spin transport processes at work. This section outlines this approach and shows how it can encompass a class of models developed to infer microstructure from large-scale measurements.

Consider a situation where two continua are present, each with its own transport process. If the compartments are spatially well-mixed, we can model this as a pair of continuity equations, each weighted by a volume fraction indicating the relative proportions of each continuum present in the system. i.e.

\begin{subequations}
    \begin{align}
        \pdd{\mathbf{M}_1(t)}{t} =& R_1\mathbf{M}_1 +\div{J_1(\mathbf{M}_1)} - E_{12}\mathbf{M_1} +E_{21}\mathbf{M}_2\\
        \pdd{\mathbf{M}_2(t)}{t} =& R_2\mathbf{M}_2 +\div{J_2(\mathbf{M}_2)} - E_{21}\mathbf{M_1} +E_{12}\mathbf{M}_2
    \end{align}
    \label{two_continua}
\end{subequations}
where $\mathbf{M}_i$ is the magnetisation in compartment $i$, $J_i$ describes the flux in a compartment and $E_{ik}$ defines the rate of exchange between compartment $i$ and compartment $k$. To ensure that spins are conserved by the exchange process we also require that
\beq
    \mathtt{Trace}(E)=1.
	\label{eq:exchng_cons}
\eeq
	
The overall magnetisation of the system is the sum of the two components
\beq
    \mathbf{M} = \mathbf{M}_1 + \mathbf{M}_2.
    \label{netmagn}
\eeq
We have assumed here that the exchange between compartments constant in space, and that the exchange process itself is describes by a Poisson process. In this form, the exchange process can be written as a matrix where
\beq
E=\left(\begin{array}{cc}
        E_{11} & E_{12} \\
        E_{21} & E_{22} \\
        \end{array}\right)
\eeq
The presence of more than one continuum introduces a new degree of freedom to the system -- a spin's presence in one or other compartment. The exchange matrix governs how spins move between the two populations. 

In the absence of exchange, the system describes a simple superposition of two non-interacting populations -- The $E$ matrix becomes diagonal and acts to define the relative weight of each contribution. In this case we can easily construct the two-tensor model, which has been employed in diffusion MRI as a model of crossing fibres. Here we assume that the relaxation parameters of both compartments are the same, i.e. that $R_1=R_2$, and that $E$ is diagonal and defines the volume fractions. This means that each equation can be solved independently using the same methodology we used to solve the tensor model, and the resulting solution is given by \eq{netmagn}, with the relative volume fractions from the $E$ matrix. From here we can see that the two-tensor model can be thought of as the solution to a pair of continua with no exchange. We can also see that this can be readily generalised to three or more compartments.

Another important case of eqs. \ref{two_continua} is the K\"arger model. Here we assume two diffusing population with the same relaxation properties with exchange. As before, we can construct this system from \eq{two_continua}. First, we set $R_1=R_2$, and make the additional assumption that diffusion in both compartments is isotropic, which reduces the tensors in each compartment to scalar diffusivities $D_1$ and $D_2$. Finally, following the methodology of Fieremans et al. \cite{FieremansKarger2010}, we assume that $E$ is defined by a constant rate matrix
\beq
    E=\left(\begin{array}{cc}
            \tau_1^{-1} & -\tau_2^{-1}\\
            -\tau_1^{-1} & \tau_2^{-1}\\
        \end{array}
    \right)
\eeq
where the matrix element are mean residence times of a spin in each compartment. An exchange process of this form assumes a Poisson process -- there is a fixed probability per unit time of an individual spin transitioning from on compartment to the other, regardless of location and independently of time. Fieremans et al note that this matrix takes the form of a rotation between compartments. The solution of the model takes the form of a familiar biexponential, but with diffusivities and volume fractions modified by the rotation caused by the exchange matrix. As before, this model can be generalised to three or more compartments, and we might also consider the case where diffusion is not isotropic, which is also soluble.

Both of these models assume that the two continua are spatially well-mixed, and neither assumes any explicit spatial arrangement. Conceptually they can be though of as models of mixed fluids, with and without exchange. 

\subsection{Other models}
So far we have shown how various models are related via a Bloch-Torrey framework with a generalised flux. Many of these models make the assumption that spin transport happens via a time-independent flux, and they all consider processes in which the environment is not explicitly modelled. Whilst this is a large class of models, there are nevertheless other models which make more general assumptions that are worth mentioning here. Some of these models may be captured by the current framework, and others not. We mention these others for the sake of completeness, and to show how different approaches to modelling the diffusion weighted signal relate to the continua discussed so far.

\subsubsection{Models with boundary conditions}
\label{with_bounds}
One research direction over the last several years has been to attempt to infer details of tissue geometry by making strong assumptions about the boundary conditions experienced by diffusing spins. These models take the multiple continua approach and impose choices of geometric boundary conditions. These are the so-called microstructure models, and include CHARMED \cite{charmed}, ActiveAx \cite{activeax}, NODDI \cite{noddi}, and VERDICT \cite{verdict}. 

Diffusion is assumed to be well-modelled by two or more sub-populations, usually without exchange, and each with strong assumptions over boundary conditions.  In addition to free space, solutions of the diffusion equation are known in which spins are restricted by surfaces such as planes, cylinders, and spheres. These solutions contain parameters describing structural size such as radius and orientation. 

The general technique is to assume a solution in each compartment and fit a weighted sum of them to a set of diffusion-weighted measurements. It is common to make a strong association between these compartments and specific structures in the tissue. CHARMED, for example, assumes a population of parallel cylinders as a model of diffusion in white matter axons and a conventional tensor model which is associated with diffusion around them in the extra cellular space. The parameters of the compartments and the weights of all contributions are then fitted to diffusion-weighted data to get estimates of structural parameters.
NODDI takes this approach one step further and constructs compartments in which intracellular diffusion is modelled by a stick (i.e. a cylinder of zero radius) convolved with a distribution of orientations. Here the approach fits the parameters of the chosen distribution to infer estimates of structural properties.

One difficulty with all of these approaches is combinatorial. As Panagiotaki et al \cite{lauraHierarchyPaper} discuss, there are a number of possibilities for choices of boundary conditions per compartment and the larger the number of compartments, the larger the number of combinations. This makes model-choice a non-trivial undertaking. The extracellular space is also potentially problematic, since these models in general make no statement about packing or explicit spatial relationships. The presence or absence of long-range order has an important effect on diffusion dynamics \cite{novikov}, but is entirely overlooked in these models. There is also the question of numerical stability. More complex models requires larger data sets to fit to, and have more complex fitting landscapes. This in turn leads to questions of convergence and uniqueness.

Finally, it is frequently claimed that these models provide quantitative estimates of tissue microstructure. Although some publications have shown that they are capable of capturing trends in microstructural properties or summary statistics (e.g. \cite{activeax}), estimates are frequently biased. These techniques work by assuming a set of boundary conditions -- there is no guarantee that the choice is representative or optimal. That is not say that such models are without merit, however. VERDICT has been shown to be a useful biomarker of the presence of tumour tissue in ex vivo prostate studies, for example, merely that care is needed when interpreting the results of these analyses -- parameter estimates cannot be taken literally.

\subsubsection{Time-dependent diffusivities}
A different theoretical approach has been taken by Novikov and Kiselev \cite{novikov}. This approach concentrates on the behaviour of the diffusion process in the presence of disordered barriers. This can be thought of an environment of randomly-oriented intersecting rods or planes, or an alternate formulation considers patches with open boundaries and different diffusivities.

A rigorous theoretic analysis shows that these assumptions lead to the idea of a time-dependent diffusivity. Diffusion gradually slows down the longer the system is observed and the rate of decay contains information about the structural properties of the environment. 

This approach is a departure from the assumptions of most other models used in diffusion MRI, and has gained traction in light of evidence from simulation \cite{saadISMRM2014}. Theoretically, the idea of a time-dependent has been extensively studied in the theoretical literature and is related to the theory of critical phenomena \cite{critical_exps}. 

Interestingly, time-dependent diffusivities are the solutions of more than one type of diffusion equation, opening up interesting theoretical possibilities for mappings between models and alternative interpretations of model parameters.

\subsubsection{Subdiffusion and space-time fractional processes}
All the models discussed so far have assumed regular evolution in time, and as such have been well-described by a first order time derivative. It is possible, however, to make different assumptions regarding diffusion dynamics
and construct alternate formulation of the Bloch-Torrey equation on this basis.

In \secn{sec:strexp} we showed how making the assumption of a random walk with step lengths distributed according to a power law lead to the stretched-exponential model. This idea can be generalised such that spins wait in position for an interval of time before taking the next step in the trajectory. The distribution of these waiting times turns out to have an important effect on the dynamics of diffusion.

Although there is often no explicit mention of waiting times when considering regular Brownian motion, the theory does make the assumption that successive steps in the trajectory are uncorrelated. This is equivalent to assuming that the process is Markovian, i.e. the current state of the system depends only on the previous state, not the entire history of the dynamics.

Waiting time distributions with well-defined finite first moments all lead to Markovian dynamics over a particular choice of timescale, so waiting time distributions given by (for example) delta functions or Poisson distributions lead to Markovian diffusion \cite{bronson}.

One class of model that breaks this assumption is the continuous time random walk (CTRW). In this case diffusing particles take steps in a trajectory with lengths drawn from a power law, such as in \secn{sec:strexp} but between steps each particle waits for a time drawn from a different power law distribution with its own exponent.

These assumptions lead to the space-time fractional diffusion equation, which has been applied to diffusion imaging by Magin and co-workers \cite{richard, carson}. This model predicts a form for the signal decay curve which is dependent on both exponents and thus estimates both exponents.

This approach employs a generalised description of diffusion itself, rather than making direct microstructural assumptions. Under this approach, diffusion dynamics evolve via a fractional time operator, similarly to the fractional space operator that arises in the stretched exponential model. The magnetisation dynamics, (i.e. the Bloch terms), however, still evolve according to a linear time operator. This duality is not compatible with the current framework, although a generalised, fractional form for the continuity equation could potential include it.

Perhaps more interestingly, the solution of a fractional diffusion equation with a constant diffusivity is formally equivalent to the solution of a regular diffusion equation with a time-dependent diffusivity, suggesting a connection between Magin approach and that of Novikov. This has yet to be explored in the literature.

\section{Discussion and conclusions}
This article has explored the Bloch-Torrey equations used to describe diffusion MRI from the point of view of the continuity equation. We have shown that if we derive the equation using a general transport process we are lead to a viewpoint in which many of the models employed in the diffusion MRI literature can be seen in a single context. We can also assess how natural each set of assumptions is, and how different choices relate to each other. We have reviewed also a large portion of the modelling literature in this context. We have also seen how thinking of the Bloch relaxation terms in the context of affine transformations and homogeneous coordinates is helpful in revealing their underlying simplicity.

We should start by noting that, of course, this framework does not capture all the subtleties of diffusion in complex environments, but we are able to see the relationships between different models of a diffusion signal that are otherwise difficult to observe. The fact that ``classical'', Gaussian diffusion leads to the diffusion tensor is well-known, but it is less well known that a L\'evy walk leads directly to a stretched exponential or that the value of the exponent interpolates between velocity-weighted phase contrast and regular diffusion-weighted contrast.


The models presented here are, of course, far from the end of the story. The generalised flux allows us to connect with the rich literature on different kinds of random walk processes (see e.g. \cite{othmer} for a review) which allow models which consider bias, correlation, and the statistics of self-avoiding walks and how this microscopic transport information might be measured from MRI data. The assumptions behind the Fractional Fickian flux, for example, may be adjusted to construct a model of fractional flow. The current framework allows for the potential to develop imaging methodologies based on novel flux assumptions not hitherto explored in the literature. We can also consider transport processes in spaces with novel topologies. The Eigenvalues of the Laplacian on binary branching networks are known \cite{laplacian_eigs_on_branching_nwks}, for example, and may provide models of flow through capillary networks and novel models of perfusion which include statistics of the network topology.

A generalisation is to consider different conserved quantities in the continuity equation. The current framework considers magnetisation in analogy with the conservation of mass, but we might also define quantities in analogy with the conservation of momentum. This leads to a Navier-Stokes-like equation and a novel class of models of imaging fluid-like properties in a sample from the MR signal. We might also consider the conservation of energy, which leads to equations with terms controlled by temperature and pressure and hence the potential for models aiming to measure these quantities from image data. 

A conservation of energy approach would also provide a closer link between the transport and relaxation physics than is present in the current treatment as energy is exchanged between spins and their environment. This is also connected to recent work such as \cite{langevin_version} which employs a Langevin equation approach to investigate a description of diffusion-weighted imaging. This more naturally links to thermodynamics.

The diffusion-MRI modelling literature often frames techniques in terms of microstructure, and whilst this is a perfectly valid viewpoint, it is not the only way to think about diffusion models. In many cases, though, the different interpretations provide complimentary viewpoints. The diffusion tensor can be viewed as an assumption that microstructure is sufficiently smoothed out that diffusion remains regular, or it can be thought of as a regular diffusion process with orientation dependence in its diffusivity. The microstructural viewpoint may be thought of as more fundamental, but provides no additional information or insights. The transport viewpoint anchors our thinking to what can be measured in the signal. Both are equally valid, and it is not necessary to assume one at the expense of the other. 

We also note that many of the solutions of the equations presented here are complex, containing both phase and frequency information in real and imaginary parts. The way in which different processes affect the each element varies with different processes, and suggests that the current standard procedure in diffusion MRI, in which frequency and phase information are combined into a signal magnitude at the start of the analysis may in fact be throwing away useful information. Modelling both the real and imaginary parts of the signal and fitting to frequency and phase information separately may be an interesting new research direction, although there would be a cost in reduced SNR. Keeping phase and magnitude separate also has the advantage of keeping the noise distribution Gaussian. Combining measurements together leads to a Rician noise distribution in the resulting data which can complicate data analysis, post-processing and fitting. The literature already contains at least one example of an approach which examines both frequency and phase information in the same model. Yin et al \cite{DMRE} have developed a scheme which measures the elastic and diffusion properties of tissue simultaneously. Elastographic information is derived from coherent motion and its effect on the signal phase, whereas diffusion information is derived from changes in the real part of the signal. 

To some readers the continuity equation-based approach will seem obvious, to others it will be unfamiliar. The approach is extremely common in the theoretical literature, although we are unaware of an explicit derivation of the continuity equation for vector-tagged particles other than in the case of momentum. The approach is a natural one under the assumption that although particle motion in the presence of field gradients causes a change in magnetisation the reverse is not true: changes in magnetisation due to field gradients do not cause a significant change to particle motion. This is not true in many circumstances (in plasmas, for example), but we believe is a reasonable approximation for particles in the biological fluids under consideration in typical MRI experiments. Indeed, as Callaghan \cite{callaghan} points out, measuring particle displacement via changes in their nuclear magnetisation is a highly decoupled procedure -- part of the reason that diffusion MRI is such as appealing modality for studying the diffusion process. As such, we believe that emphasising spin flux and the resulting generalisation of the Bloch-Torrey equation is a useful and applicable concept. 

\bibliography{lintrans}{}
\bibliographystyle{plain}

\section*{Supplementary materials}
\subsection*{Solution of the Bloch-Torrey equation in the lab frame}
We've seen that we can do this in the rotating frame and obtain the familiar Bloch relaxation terms. This repeats that analysis with the form of the Bloch matrix which we includes the precessional terms. We start with the Bloch equation including the precessional terms.
\beq
    \pdd{\mathbf{M}}{t} = R\mathbf{M}
\eeq
where
\beq
    R=\left(\begin{array}{cccc}
             -\frac{1}{T_2} & -\gamma B_z & 0 & 0 \\
             \gamma B_z & -\frac{1}{T_2} & 0 & 0 \\
             0 & 0 & -\frac{1}{T_1} & \frac{M_0}{T_1} \\
             0 & 0 & 0 & 1\\
     \end{array}\right)
     \label{eq:Blochmatlab}
\eeq
As before, we know that the solution of the Bloch equation is given by \eq{eq:vecdiffsoln} and that all we need to do is construct the matrix exponential of $R$ as defined in \eq{eq:matexp}. Diagonalising \eq{eq:Blochmatlab} gives the set of Eigenvectors
\beq
    R_V=\left(\begin{array}{cccc}
            0 & 0 & i & -i\\
            0 & 0 & 1 & 1\\
            M_0 & 1 & 0 & 0\\
            1 & 0 & 0 & 0\\
    \end{array}\right)
    \label{eq:labevecs}
\eeq
which has the inverse
\beq
    R_V^{-1}=\left(\begin{array}{cccc}
            0 & 0 & 0 & 1\\
            0 & 0 & 1 & -M_0\\
            \frac{i}{2} & \frac{1}{2} & 0 & 0\\
           -\frac{i}{2} & \frac{1}{2} & 0 & 0\\
    \end{array}\right).
    \label{eq:invlabevecs}
\eeq
$R$ also has eigenvalues given by
\beq
    R_e=\left(\begin{array}{cccc}
        0 & 0 & 0 & 0 \\
        0 & -\frac{1}{T_1} & 0 & 0 \\
        0 & 0 & -\frac{1}{T_2} -i\gamma B_z& 0 \\
        0 & 0 & 0 & -\frac{1}{T_2} +i\gamma B_z\\
    \end{array}\right).
    \label{eq:labtotalevals}
\eeq
Which we note is the sum of two diagonal matrices
\beq
R_e = R_e^\gamma + R_e^T = \left(\begin{array}{cccc}
                                0 & 0 & 0 & 0 \\
                                0 & 0 & 0 & 0 \\
                                0 & 0 & -i\gamma B_z & 0 \\
                                0 & 0 & 0 & i\gamma B_z \\
                            \end{array}\right)
                            +
                             \left(\begin{array}{cccc}
                                0 & 0 & 0 & 0 \\
                                0 & -\frac{1}{T_1} & 0 & 0 \\
                                0 & 0 & -\frac{1}{T_2} & 0 \\
                                0 & 0 & 0 & -\frac{1}{T_2} \\
                            \end{array}\right).
\eeq
Next we note that
\beq
    e^{Rt} = R_Ve^{R_et}R_V^{-1} = R_Ve^{(R_e^\gamma + R_e^T)t}R_V^{-1}
\eeq
and furthermore that
\beq
    e^{Rt} = R_Ve^{R_e^\gamma t}e^{R_e^Tt}R_V^{-1}
\eeq
which means
\beq
    e^{Rt} = R_Ve^{R_e^\gamma t}R_V^{-1}R_Ve^{R_e^Tt}R_V^{-1}
\eeq
because $R_V^{-1}R_V = I$, the identity matrix. 

This is the product of two matrix exponentials: one governing relaxation and the other precession. These can be explicitly evaluated by performing each multiplication. The relaxation matrix is
\beq
    R_Ve^{R_e^Tt}R_V^{-1} = \left(\begin{array}{cccc}
                                e^{-\frac{t}{T_2}} & 0 & 0 & 0 \\
                                0 & e^{-\frac{t}{T_2}} & 0 & 0 \\
                                0 & 0 & e^{-\frac{t}{T_1}} & -M_0 + e^{-\frac{t}{T_1}}M_0\\
                                0 & 0 & 0 & 1 \\
                            \end{array}\right)
                            \label{eq:labrelaxterms}
\eeq
which is equivalent to the solution in the rotating frame. In the lab frame we also have a contribution from the precessional terms given by
\beq
    R_Ve^{R_e^\gamma t}R_V^{-1}= \left(\begin{array}{cccc}
            \frac{1}{2}\left(e^{-i\gamma B_zt} + e^{i\gamma B_zt}\right) & \frac{i}{2}\left(e^{-i\gamma B_zt}-e^{i\gamma B_zt}\right) & 0 & 0\\
            -\frac{i}{2}\left(e^{-i\gamma B_zt}-e^{i\gamma B_zt}\right) & \frac{1}{2}\left(e^{-i\gamma B_zt} + e^{i\gamma B_zt}\right) & 0 & 0\\
                                0 & 0 & 1 & 0\\
                                0 & 0 & 0 & 1\\
                                \end{array}\right).
                                \label{eq:labrotterms}
\eeq
Finally we note that
\begin{subequations}
    \begin{align}
        \cos\theta &= \frac{1}{2}\left(e^{-i\theta} + e^{i\theta}\right)\\
        \sin\theta &= \frac{i}{2}\left(e^{-i\theta} - e^{i\theta}\right)
    \end{align}
    \label{trigIds}
\end{subequations}
which means \eq{eq:labrotterms} can be rewritten as
\beq
     R_Ve^{R_e^\gamma t}R_V^{-1}= \left(\begin{array}{cccc}
                                        \cos(\gamma B_zt) & \sin(\gamma B_zt) & 0 & 0\\
                                        -\sin(\gamma B_zt) & \cos(\gamma B_zt) & 0 & 0\\
                                        0 & 0 & 1 & 0\\
                                        0 & 0 & 0 & 1\\
                                    \end{array}\right)
\eeq
which is a rotation matrix about the $z$ axis through an angle $\gamma B_zt$, which is the precessional motion due to a static field parallel to the $z$-axis. Hence we can see that the choice of frame acts only to introduce a multiplicative rotation matrix to the relaxation terms.

\subsection*{Fractional version of Fick's law}
We assume a L\`evy walk-type process, which is fractional in space but not time. We start with the distribution of step lengths
\beq
    P(\dx\ge\Delta X) = \frac{1}{\Gamma (1-\beta)}(\dx)^{-\beta}
\eeq
The derivative with respect to $\dx$ gives the probability density function,
\beq
    p(\dx) = \frac{\beta}{\Gamma(1-\beta)}(\dx)^{-\beta-1}.
\eeq
where in both cases the gamma function is a necessary normalising factor \cite{Meerschaert2003}.

We can also think of the displacement $\dx$ as the difference between a location $x$ and a starting point $x_0$, i.e.
\beq
    P(x-x_0)\ge\Delta X)= \frac{1}{\Gamma (1-\beta)}(x-x_0)^{-\beta}
    \label{disp_prob}
\eeq
and
\beq
    p(x-x_0)= \frac{\beta}{\Gamma(1-\beta)}(x-x_0)^{-\beta-1}.
    \label{disp_pdf}
\eeq
We note that the displacement probability is a function only of the step length, and has nothing to say about whether spin displacement is in the positive or negative sense, we assume that spins are displaced in either direction with equal probability. Since we are only interested in this in one direction we need to multiple both by a half. 

We can define the total flux through a point $x$ as the transport from all other points at least as far as $x$, i.e.
\beq
    J(\mathbf{M})= \frac{1}{2}D\left[\int_{-\infty}^xP(x-x_0)\mathbf{M}(x_0)dx_0 + \int_x^{\infty}P(x_0-x)\mathbf{M}(x_0)dx_0\right].
    \label{flux1}
\eeq
By substitution from eq.-\ref{disp_prob}, the first term in the above becomes
\beq
    \int_{-\infty}^xP(x-x_0)\mathbf{M}(x_0)dx_0 = \frac{1}{\Gamma(1-\beta)}\int_{-\infty}^x(x-x_0)^{-\beta}\mathbf{M}(x_0)dx_0.
    \label{lhs_int_first}
\eeq
which is the definition of the RHS (or positive-$x$) Weyl fractional derivative of $\mathbf{M}$ of order $\beta$ with respect to $x$ \cite{tomfractionalbook}. We denote this operator $^W\!\!D_{x+}^\beta$. 

Similarly, substituting eq.-\ref{disp_prob} into the second term on the RHS of of eq.-\ref{flux1} gives
\beq
    \int_x^{\infty}P(x_0-x)\mathbf{M}(x_0)dx_0 = \frac{1}{\Gamma(1-\beta)}\int_x^{\infty}(x_0-x)^{-\beta}\mathbf{M}(x_0)dx_0,
    \label{rhs_int}
\eeq
which is the definition of the LHS (or negative) Weyl derivative, denoted by  $^W\!\!D_{x-}^\beta$.

Combining these back into eq.\ref{flux1} gives us
\beq
    J(\mathbf{M}) = -\frac{1}{2}D\left[ ^W\!\!D_{x+}^\beta + ^W\!\!\!D_{x-}^\beta\right]\mathbf{M}.
    \label{flux2}
\eeq
We now note that the Reisz-Weyl fractional derivative can be defined as
\beq
    \frac{\partial^\beta}{\partial|x|^\beta} = -\frac{1}{2}\left[ ^W\!\!D_{x+}^\beta + ^W\!\!\!D_{x-}^\beta\right]
    \label{reisz}
\eeq
which means eq.-\ref{flux2} becomes
\beq
    J(\mathbf{M})=-D\frac{\partial^\beta}{\partial|x|^\beta}\mathbf{M}
    \label{flux3}
\eeq
which is the fractional form of Fick's law appropriate to particles executing random walks with step lengths drawn from a power law distribution.

Finally, we note that the Reisz-Weyl derivative has a well defined Fourier transform \cite{tomfractionalbook} which means that assuming $D$ is constant in space, the Fourier transform of the fractional flux is given by
\beq
    FT\left[J(\mathbf{M})\right] = -D|q|^\beta \mathbf{m},
    \label{eq:ft_frac_flux}
\eeq
where $\mathbf{q}$ is the Fourier variable and $\mathbf{m}$ is the Fourier Transform of the magnetisation.

\end{document}